\documentclass[a4paper,fleqn,usenatbib]{mnras}

\usepackage[T1]{fontenc}
\usepackage{ae,aecompl}

\usepackage{graphicx}	
\usepackage{amsmath}	
\usepackage{amssymb}	

\title[Subarcsecond MIR emission in nearby QSO]{The dusty tori of nearby QSOs as constrained by high-resolution mid-IR observations}

\author[M. Mart\'inez-Paredes et al.]{
M. Mart\'inez-Paredes,$^{1,2}$\thanks{E-mail: m.martinez@crya.unam.mx (M-MP)}
I. Aretxaga,$^{1}$
A. Alonso-Herrero,$^{3,4,5}$
O. Gonz\'alez-Mart\'in,$^{2}$
\newauthor
E. Lop\'ez-Rodr\'iguez,$^{16,17,18}$
C. Ramos Almeida,$^{6,14}$
A. Asensio Ramos,$^{6,14}$
\newauthor
T. Diaz Santos,$^{8}$
M. Elitzur,$^{12,13}$ 
P. Esquej,$^{3}$
A. Hern\'an-Caballero,$^{19}$
K. Ichikawa,$^{5,15}$
\newauthor
R. Nikutta,$^{10}$ 
C. Packham,$^{5}$
M. Pereira-Santaella$^{11}$ and 
 C. Telesco$^{9}$ 
\newauthor 
\\
$^{1}$Instituto Nacional de Astrof\'{\i}sica, \'Optica y Electr\'onica (INAOE), Luis Enrique Erro 1, Sta. Ma. Tonantzintla, Puebla, Mexico\\
$^{2}$Instituto de Radioastronom\'{\i}a y Astrof\'{\i}sica UNAM, Apartado
Postal 3-72 (Xangari), 58089 Morelia, Michoac\'an, Mexico\\
$^{3}$Centro de Astrobiolog\'ia (CAB, CSIC-INTA), ESAC Campus, E-28692 Villanueva de la Ca{\~n}ada, Madrid, Spain\\
$^{4}$Department of Physics, University of Oxford, Oxford OXI 3RH, UK\\
$^{5}$Department of Physics and Astronomy, University of Texas at San Antonio, San Antonio, TX 78249, USA\\
$^{6}$Instituto de Astrof\'{\i}sica de Canarias (IAC), E-38205 La Laguna,
Tenerife, Spain\\
$^{8}$N\'ucleo de Astronom\'ia de la Facultad de Ingener\'ia, Universidad Diego Portales, Av. Ej\'ercito Libertador 441, Santiago, Chile\\
$^{9}$Department of Astronomy, University of Florida, Gainesville, FL 32611, USA\\
$^{10}$National Optical Astronomy Observatory, 950 N Cherry Ave, Tucson, AZ 85719, USA\\
$^{11}$Department of Physics, University of Oxford, Keble Road, Oxford OX1 3RH, UK\\
$^{12}$Astronomy Dept., University of California, Berkeley, CA 94720, USA\\
$^{13}$Physics $\&$ Astronomy, University of Kentucky, Lexington, KY 40506, USA\\
$^{14}$Departamento de Astrofísica, Universidad de La Laguna, E-38206, Tenerife, Spain\\
$^{15}$National Astronomical Observatory of Japan, 2-21-1 Osawa, Mitaka, Tokyo 181-8588, Japan\\
$^{16}$SOFIA/USRA, NASA Ames Research Center, Moffett Field, CA 94035, USA\\
$^{17}$Department of Astronomy, University of Texas at Austin, 1 University Station C1400, Austin, TX 78712, USA\\
$^{18}$McDonald Observatory, University of Texas at Austin, Austin, TX 78712, USA\\
$^{19}Departamento de Astrof\'isica, Facultad de CC. F\'isicas, Universidad Complutense de Madrid, E-28040 Madrid, Spain$}

\date{Accepted XXX. Received YYY; in original form ZZZ}

\pubyear{2016}

\begin{document}

\label{firstpage}

\date{Accepted ---. Received ---; in original form ---}

\pagerange{\pageref{firstpage}--\pageref{lastpage}}
\maketitle

\begin{abstract}
We present mid-infrared (MIR, $7.5-13.5\mu$m) imaging and spectroscopy
observations obtained with the CanariCam (CC) instrument on the
10.4m Gran Telescopio CANARIAS for a sample of 20 nearby, MIR bright
and X-ray luminous QSOs. We find that for the majority of QSOs the MIR
emission is unresolved at angular scales $\sim 0.3$~arcsec,
corresponding to physical scales $\lesssim600$~pc.
We find that the higher-spatial resolution CC spectra have similar
shapes to those obtained with \emph{Spitzer/IRS}, and hence we can 
assume that the spectra are not heavily contaminated by extended emission 
in the host galaxy.
We thus take advantage of the higher signal to noise 
\emph{Spitzer/IRS} spectra, as a fair representation of the nuclear emission, 
to decompose it into a 
  combination of active galactic nuclei (AGN), polycyclic aromatic 
hydrocarbon (PAH) and 
stellar components. In most cases the
AGN is the dominant component, with a median contribution 
of 85 per cent of the continuum light at MIR (5-15 $\mu$m) within the IRS slit.  This
IR AGN emission is well reproduced by {\sc clumpy} torus models.  We
find evidence for significant differences in 
the parameters that describe
the dusty tori of QSOs when compared with the same parameters of 
Seyfert 1 and 2 nuclei. In particular, we find a lower
number of clouds ($N_{0}\lesssim12$), steeper radial distribution of clouds
($q\sim1.5-3.0$), and clouds that are less optically thick
($\tau_{V}\lesssim100$) than in Seyfert 1, which could be
attributed to dusty structures that have been partially evaporated and
piled up by the higher radiation field in QSOs. We find that the
combination of the angular width $\sigma_{torus}$, viewing angle $i$,
and number of clouds along the equatorial line $N_{0}$, produces large
escape probabilities ($P_{esc}>2$ per cent) and low geometrical covering
factors ($f_{2}\lesssim0.6$), as expected for AGN with broad lines in their optical spectra.  
\end{abstract}




\section{Introduction}
The dusty torus \citep[e.g.,][]{Rowan-Robinson77, Krolik_Begelman88}
is the cornerstone of the unified scheme for Active Galactic Nuclei
\citep[AGN, e.g.,][]{Antonucci93, UrryPad95}. This framework
attributes the differences between type 1 and type 2 AGN to the
orientation of a putative dusty torus that surrounds the central
engine around the supermassive black hole. Type 1 AGN show broad
permitted optical emission lines (with full-width half-maxima
$FWHM\sim10^{3}-10^{4}$ km s$^{-1}$) and narrow permitted and
forbidden emission lines ($FWHM\sim500$ km s$^{-1}$), while type 2 AGN
only show permitted and forbidden narrow emission lines
($FWHM\sim400-500$ km s$^{-1}$), as the broad line region is obscured
by the torus under this framework. The torus absorbs the emission of
the central engine and re-radiates it in the mid-infrared (MIR), such
that at $\sim5-35$ $\mu$m it is the dominant component
\citep[e.g.,][]{UrryPad95, Urry03, Packham05, Radomski08}.  Hence the
shape of the MIR spectral energy distribution (SED) depends crucially
on the configuration, providing a clean insight into its geometry and
composition. If the dust is homogeneously distributed in the torus,
the IR emission that arises from the inner region of the torus (hot
and optically thin region) should be larger than the emission observed
through the torus, which is optically thick, resulting in a steeper
SED \citep[e.g.,][]{PierKrolik92, GranatoDanese94,
  EfstathiouRowan-Robinson95}. On the other hand, if the dust is in a
clumpy distribution of optically thick clouds that do not fill all the
volume, then the dependence of the luminosity with the viewing angle
decreases and the SED becomes flatter
\citep[e.g.,][]{2008ApJ...685..147N, 2008ApJ...685..160N,
  Stalevski12}. However, \citet[][]{Feltre12}, in a detailed
  comparison of smooth and clumpy models, find that both configurations
  can predict similar MIR continuum shapes for different
  model parameters, but their predicted NIR slopes are different.

MIR observations acquired with 8m-class ground-based telescopes
provide high spatial resolution data ($\lesssim 0.3$ arcsec) crucial
to isolate the emission of the dusty torus and the AGN from its host
\citep[e.g.,][]{Krabbe01, Horst06, Mason06, Horst08, Gandhi09,
  Levenson09}. During the last decades these MIR observations have
constrained the spatial extension of tori in nearby Seyfert galaxies
to be $\lesssim 5$ pc \citep[e.g.,][]{Jaffe04, Packham05, Tristam07,
  Radomski08}, giving support to models where the torus is fragmented
into clouds that form a clumpy obscuring medium.

 One of the largest
sample of Seyfert galaxies studied so far with ground-based MIR
observations suggests that their classification as type 1 or 2 does
not only depend on the viewing angle but also on the intrinsic
geometry of the cloud distribution \citep[e.g.,][]{Ramos09,
  2011ApJ...731...92R, 2011ApJ...736...82A, Mateos16}. In a previous
work, \citet{2015ApJ...803...57I} studied a sample of type 1 and 2
Seyfert galaxies, with and without the signs of broad polarized lines,
and found that the intrinsic properties of the tori are also
intrinsically different. In a recent study, \citet{Garcia-Burillo16}
modeled the torus in the nearby Seyfert galaxy NGC1068 using ALMA plus
nuclear NIR and MIR data. They found that the nuclear emission at
submilimeter wavelengths (432 $\mu$m) is consistent with a clumpy
distribution of the dust.

\begin{figure*} 
	\centering 
	\begin{minipage}{1. \textwidth} 
		\includegraphics[width=1.\columnwidth]{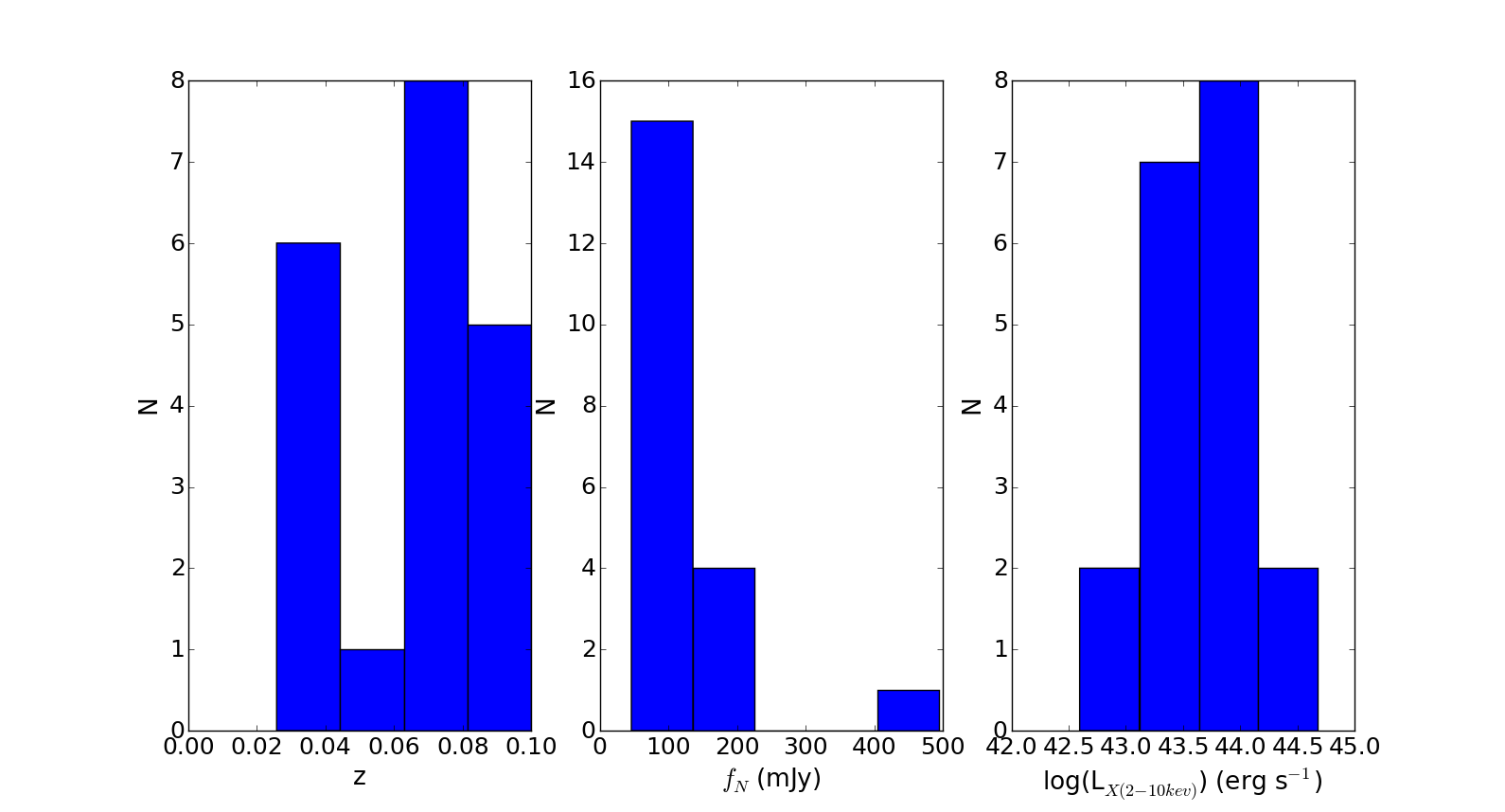} 
		\caption{From left to right, distributions of redshift, 
$N$-band flux-density, and $2-10$ keV X-ray luminosity for the QSOs in our sample.}     
		\label{fig1} 
	\end{minipage} 
\end{figure*}

While much effort has been devoted to characterise the dusty torus of
Seyfert galaxies using high angular resolution data
\citep[e.g.,][]{2011ApJ...731...92R, 2011ApJ...736...82A,
  2015ApJ...803...57I, 2010A&A...515A..23H} and low angular resolution
data \citep[e.g,][]{Lutz04, 2006ApJ...649...79S, Ramos07, Ichikawa12,
  Gonzalez15}, the study of the dusty torus in Quasi Stellar Objects
(QSOs) has been limited to low resolution data \citep[$\sim3$~arcsec,
  e.g.,][]{2009ApJ...705..298N, 2009ApJ...707.1550N, Mateos16, Ichikawa17}. This
is mainly due to their compactness and sparseness in the local
Universe. The point-like morphology of QSOs does not allow us to
disentangle the extended emissions from the host galaxy in the
immediate vicinity of the central engine, which can be an important
contaminant that impacts greatly the results on the properties of the
torus. Therefore, MIR high angular resolution observations offer a
good opportunity to step forward in their understanding.

Previous studies in the IR reveal that the majority of Palomar-Green
(PG) QSOs \citep{Green86} present signs of a recent galactic
interaction \citep[e.g,][]{2009ApJS...182..628V} and show polycyclic
aromatic hydrocarbon (PAHs) emission in the \emph{Spitzer/IRS}
spectrum \citep[][]{2006ApJ...649...79S} indicating the presence of
starbursts at scales of a few kpc for the nearest objects
($z<0.1$). Moreover, it is also well known that these objects present
prominent 10 $\mu$m silicate emission features, suggesting the
presence of a dusty torus.  Although, it is not entirely clear whether
all this emission actually arises from the inner regions of a face-on
torus or whether part of it comes from an extended silicate region
towards the Narrow Line Region (NLR) \citep[][]{Netzer08}. The study
of the IR emission in this class of AGN requires a complex combination
of components (e.g., a combination of the torus, NLR and/or
starburst). Some works have also included an additional component of
hot dust emission from the inner region of the torus in order to find
a successful SED fitting of large aperture observations, especially
between 1 and 8 $\mu$m \citep[e.g.,][]{Schweitzer08, Mor_Netzer,
  2010A&A...515A..23H}.

In this work we present MIR high angular (0.3-0.4 arcsec) resolution
imaging and spectroscopy obtained with CanariCam
\citep[CC,][]{2003SPIE.4841..913T, Packham05} in the 10.4m Gran
Telescopio CANARIAS (GTC) of a sample of 20 nearby QSOs, with the aim
to constrain the geometrical parameters of their dusty tori and
compare them with those found in Seyfert galaxies that have been
studied with similar techniques. In Section \ref{sample} we present
the QSO sample. In Section \ref{data1} we show our new high angular
resolution observations and in \ref{ancillary} the ancillary data we
will use in the modelling.  Section \ref{analysis} presents some basic
analysis of the data, in Section \ref{results1} we present the
spectral decomposition into AGN and starburst components, and in
Section \ref{results2} we perform spectral fitting of the unresolved
IR SED and MIR spectroscopy using {\sc clumpy} models. We discuss the
main results in Section \ref{discussion} and our conclusions are given
in Section \ref{conclusions}. We adopt a cosmology with $H_{0}=73$ km
s$^{-1}$Mpc$^{-1}$, $\Omega_{M}=0.27$ and $\Omega_{\Lambda}=0.73$.

\section{The sample}
\label{sample}

\begin{table*}
	\begin{minipage}{1.\textwidth}
		\caption{QSO sample. Column 1 gives the name and classification as type 1, 2 or narrow-line type 1 (NL1) AGN, 
column 2 and 3 the coordinates in right ascension and declination, columns 4, 5 and 6 the
redshift, angular scale and the comoving radial distance, column 7 the absolute 
magnitude, column 8 the radio classification as radio-quiet (Q) or radio-loud (L) AGN, and column 9 the hard X-ray (2-10 kev) luminosity. \label{tab1}}

\centering
\begin{tabular}{lcccccccc}
				\hline\\
				Name (AGN-type)            &	RA (2000)	   &	DEC (2000)	&	$z^{a}$	& Scale    &d	&$M^{b}_{V}$   & Radio$^{c}$ &$L^{e}_{X (2-10 {\rm keV})}$   \\
				       & (h:m:s) & (h:m:s) &  & (kpc/arcsec)& (Mpc) & (mag) &   &   (erg s$^{-1}$)  \\
				\hline
				\hline
PG 0003+199/Mrk 335 (1) & 00:06:19.5 & 20:12:10	& 0.03 &	0.498 & 122& -19.0	& Q$^{d1}$	& 1.$9\times10^{43}$ 	\\
PG 0007+106/Mrk 1501 (1.2)& 00:10:31.0 & 10:58:30	& 0.09  & 1.602 & 363& -22.9 &	L$^{d2}$& 1.$4^{e1}\times10^{44}$  	 \\
				PG 0050+124/IZw1  (1)  & 00:53:34.9 & 12:41:36	& 0.06   &1.094  & 243 & -22.7	&	Q & 7.$1\times10^{43}$  	\\
				PG 0804+761	(1)        & 08:10:58.6 & 76:02:43	& 0.10   &1.772  & 402 & -23.2	& Q & 2.$9\times10^{44}$  \\
				PG 0844+349	(1)        & 08:47:42.4 & 34:45:04	& 0.06    &1.182  & 243 & -22.4	&	Q & 5.$5\times10^{43}$  \\
				PG 0923+129/Mrk 705 (1.2) & 09:26:03.3 & 12:44:04 & 0.03 &0.562	& 122 & -20.9	& Q$^{d3}$	& 2.$6^{e2}\times10^{43}$     \\
				PG 1211+143 (NL1)	        & 12:14:17.7 & 14:03:13	& 0.08 &1.465	& 323 & -23.3	& Q & 5.$0\times10^{43}$  \\
				PG 1229+204/Mrk 771 (1)	& 12:32:03.6 & 20:09:29	& 0.06   & 1.165 & 243& -20.0	&	Q & 3.$1\times10^{43}$    \\
				PG 1351+640	(1.5)        & 13:53:15.7 & 63:45:46	& 0.09  & 1.584  & 363& -23.3	&	Q & 1.$2^{e3}\times10^{43}$   \\
				PG 1411+442 (1)	        & 14:13:48.3 & 44:00:14	& 0.09    & 1.607  & 363& -23.9	& Q & 2.$5\times10^{43}$   \\
				PG 1426+015/Mrk 1383 (1)& 14:29:06.6 & 01:17:06	& 0.09     & 1.558   & 363 & -22.7	&	Q & 1.$3\times10^{44}$   \\
				PG 1440+356/Mrk 478 (NL1)	& 14:42:07.4 & 35:26:23	& 0.08 & 1.436	& 323 & -22.7	& Q & 5.$8\times10^{43}$   \\
				PG 1448+273	(NL1)        & 14:51:08.8 & 27:09:27	& 0.07  & 1.199	& 283& -22.2	& Q$^{d4}$	& 2.$0\times10^{43}$    \\
				PG 1501+106/Mrk 841 (1.5)	& 15:04:01.2 & 10:26:16	& 0.04 & 0.694 & 163	& -20.9	& Q$^{d5}$	& 7.$8\times10^{43}$ \\				
				PG 1534+580/Mrk 290 (1.5) & 15:35:52.3 & 57:54:09 & 0.03 & 0.569 &122	& -18.5	& Q$^{d1}$	& 1.$8\times10^{43}$	\\
				PG 1535+547/Mrk 486 (1) & 15:36:38.3 & 54:33:33 & 0.04 & 0.740 & 163 & -20.8	& --	& 4.$0^{e4}\times10^{42}$  	\\
				PG 2130+099/IIZw136 (1.5) 	& 21:32:27.8 & 10:08:19	& 0.06 & 1.165	& 243& -18.5	& Q & 3.$2\times10^{43}$  \\
				PG 2214+139/Mrk 304 (1)	& 22:17:12.2 & 14:14:21	& 0.07    & 1.213  &283  & -22.3	& Q & 6.$6\times10^{43}$    \\
			         Mrk 509	    (1.5)        & 20:44:09.7 &-10:43:25	& 0.03 & 0.657	 &122   & -22.5	& Q$^{d4}$	& 4.$8\times10^{44}$    \\	
				MR 2251-178	(1.5)        & 22:54:05.9 &-17:34:55	& 0.06 & 1.182	 &243 & -22.2	& --	& 2.$9\times10^{44}$   \\		
				\hline
			\end{tabular}\\
			{\bf References}.$^{a}$NED, $^{b}$\citet{2010A&A...518A..10V}, $^{c}$\citet{Kellerman1994}, $^{d1}$\citet{Zhou2010}, $^{d2}$\citet{Laurent-Muehleisen1997}, $^{d3}$\citet{Bicay1995}, $^{d4}$\citet{Barvainis1996}, $^{d5}$\citet{Edelson1987},$^{e}$\citet{Zhou2010}, 
$^{e1}$\citet{Piconcelli2005}, $^{e2}$\citet{Shu10}, $^{e3}$\cite{Bianchi2009}, $^{e4}$\citet{Gallo06}.
			
		\end{minipage}
	\end{table*}

We select a representative sample of 20 X-ray luminous and MIR-bright
nearby QSOs from the latest
version of the \citet{2010A&A...518A..10V} catalogue and the NASA/IPAC
Extragalactic Database\footnote{\url{https://ned.ipac.caltech.edu/}} (NED),
that meet the following criteria:  
1) redshift $z<0.1$
to obtain a minimum spatial scale of $\rm{\sim}$1.8~kpc/arcsec, so that 
for our projected 
nearly difraction-limited observations ($\lesssim 0.3$~arcsec) we sample 
spatial scales $\lesssim 600$~pc; 
2) flux density at $N$ band $f_N>0.02$ Jy to be able
to detect them with CanariCam (CC) on the 10.4~m GTC; and 3) intrinsic X-ray
luminosity $L_{X (2-10 {\rm keV})}>10^{43}$ erg s$^{-1}$,  
to focus on the most powerful AGN. 

All objects in our sample, except Mrk~509 and
MR~2251-178, are also part of the Bright Quasar Survey
\citep{1983ApJ...269..352S} and Palomar Green survey
\citep{1987ApJS...63..615N}. Mrk~509 is usually classified as a type~1 Seyfert
nucleus. However, this object fits all our selection criteria and has an
absolute magnitude ($M_{B}=-22.5$) consistent in the optical with the QSO
definition \citep[][]{2010A&A...518A..10V}. 
MR~2251-178 has not been widely studied but also fits
our selection criteria, and its X-ray luminosity places it among the most 
powerful AGN with $M_{B}=-22.2$
\citep[][]{2010A&A...518A..10V}. Note that all the
objects in our sample have \emph{Spitzer/IRS}  spectra, except
for MR~2251-178. 

The redshift, $N$-band flux density and X-ray luminosity 
distributions for the
sample are shown in Figure \ref{fig1}, and the list of 
QSOs and their literature-compiled properties are shown in Table \ref{tab1}.

\section{Observations and data reduction} 
\label{data1}
\subsection{GTC/CC 8.7 $\mu$m imaging}
\label{data}

\begin{table*}
	\begin{minipage}{1.\textwidth}
		\caption{Log of GTC/CC imaging observations in Si2 band ($8.7\, \mu$m).
 Column 1 gives the name of the QSO, column 2 the date of observations, column 3 the on-source time integration, column 4 the name of the standard star \citep{1999AJ....117.2428C}, column 5 the time elapsed between standard star and 
science target acquisition, column 6 the FWHM of the standard star, columns 7 and 8 the airmass and precipitable water vapor during the observations, column 9 the position angle that indicates the orientation of the detector on the sky, and column 10 the programme for which the data 
were acquired, where GT stands for Guaranteed Time, ESO-GTC for European Southern Observatory-GTC large programme, and MEX for open Mexican time.}
        \end{minipage}
        
        \begin{minipage}{1.\textwidth}
          \label{tab2}
		\centering
\begin{tabular}{lccccccccc} 
				\hline\hline 
				Name	& Date  &$t_{on}$ & STD & $t_{offset}$    & $FWHM_{STD}$  & Airmass & PWV& PA & Programme \\
				&  & (s)   & (s)    & (min)& (arcsec)     &       &  (mm)   &  (deg)    &   \\
				\hline
				\hline
				PG 0003+199 & 2013.09.15 & 904  & HD 2436 & 8& 0.27 & 1.18 & 11.3& 0  & GT\\
				PG 0007+106 & 2013.09.27 & 973  & HD 2436 &9 & 0.30  & 1.20& 13.6-14.0   & 0   & MEX \\
				PG 0050+124        & 2013.09.14 & 904  & HD 2436 & 9 &0.35 & 1.3  & 9.8-10.3 &  0  & GT \\
				PG 0804+761 & 2014.01.03 & $3\times209$ & HD 64307 & 42 & 0.33  & 1.47 & 9.4 & 360 & ESO-GTC\\
				PG 0844+349 & 2014.01.06 & $3\times216$ & HD 81146 & 17 & 0.34   &  1.05 & 7.3-7.1 & 360   & ESO-GTC \\
				PG 0923+129  & 2015.04.03& 695    & HD 82381 & 9 & 0.34 &   1.04 & 5.0-5.5 & 0  & MEX \\
				PG 1211+143  & 2014.03.14 & $3\times209$ & HD 107328 & 26 & 0.31  & 1.04&$<10$ & 0  &  ESO-GTC\\
				PG 1229+204  & 2014.06.08 & 1251  & HD 111067 & 8 & 0.27  & 1.04 & 6.3-6.4& 360  &  ESO-GTC\\
				PG 1351+640  & 2014.05.20& 1112  & HD114326   & 8 & 0.28  & 1.08   & 4.6& 360  & MEX \\
				PG 1411+442  & 2014.03.16& $2\times209$   & HD128902 & 2 & 0.27 & 0.96    & 4.0-3.5 & 360 & ESO-GTC \\
				PG 1426+015     & 2012.03.09& $3\times220$  &HD126927 & 58& 0.40 & 1.15 & 3.0  & 0  & ESO-GTC  \\
				PG 1440+356      & 2014.03.16& 209   & HD128902 & 4 & 0.25  & 1.13 & 4.8  & 0  &  ESO-GTC\\
				PG 1448+273  & 2014.06.08& 1112    &  HD138265 & 11 & 0.27  & 1.35   & 8.1-5.1& 360   & MEX  \\
				PG 1501+106  & 2013.08.30& $3\times209$ & HD 140573 & 7 & 0.27 &     1.48 & 5.3 & 360 &  ESO-GTC\\
				PG 1534+580 & 2015.04.04 & 695   & HD 138265  & 6 & 0.33 &  1.28 & 1.8-2.5 & 360 & MEX  \\
				PG 1535+547  & 2014.05.16& 1112    & HD138265  & 6 & 0.30    & 1.61  & 4.8-5.0& 360   & MEX \\
				PG 2130+099      & 2014.06.10& 904    &  HD206445 & 8 & 0.28   &  1.13& 7.4-7.5  & 0   & GT \\ 
				PG 2214+139$^{a}$ &  2013.09.17  & 1042    & HD220363 &  13  & 0.8   & 1.32  &  9.0 & 0 &  MEX\\
				MR2251-178   & 2013.09.17& 1043   & HD220363 & 7& 0.3 & 1.30 &  $<10$ & 0   &  MEX\\
				\hline
			\end{tabular}\\
			Notes.--$^{a}$ Bad quality image. 

		\end{minipage}
		
	\end{table*}

Twelve  QSOs in our sample were observed within the European Southern 
Observatory-GTC 
(PI: A. Alonso-Herrero, ID: 182.B-2005) and the CC Guaranteed Time 
(PI: C. Telesco) large programmes, which also include a large sample of 
other AGN with X-ray luminosities between 
$L_{2-10 \, keV}\sim3\times10^{38}-3\times10^{45}$ erg s$^{-1}$ 
\citep{2016MNRAS.455..563A}. The other 8 objects were observed with  Mexican 
time on this facility (PIs: I. Aretxaga and M. Mart\'inez-Paredes). 

All QSOs in the sample, except for Mrk~509, were observed with GTC/CC 
in imaging mode with the Si2 filter ($\lambda_{c}=8.7$ $\mu$m, $\Delta\lambda=1.1$ $\mu$m) between
March 2012 and April 2015. Note that we excluded Mrk~509 because it
has unresolved emission at MIR, as reported by
\citet{2010A&A...515A..23H}. On average, the images were acquired with
a precipitable water vapour (PWV) $\rm{\sim}$7.1 mm and typical air
mass of $\rm{\sim}$1.2. The log of the imaging observations is compiled in
Table \ref{tab1}.

In order to flux-calibrate and estimate the image quality, a standard
star was imaged with the same filter just before or after the science
target.  Considering that the theoretical
diffraction-limited 
FWHM\footnote{\url{http://www.gtc.iac.es/instruments/canaricam/canaricam.php\#Imaging}} of the point spread function
(PSF) of CC in the Si2 band is 0.19 arcsec, the
majority of QSOs have good image quality ($<FWHM>\sim0.3$ arcsec),
except for PG~2214+139, which was observed with  $FWHM\sim0.8$ arcsec.

We use the CC pipeline {\sc RedCan} developed by
\cite{2013A&A...553A..35G} for the reduction and analysis of
ground-based MIR CC and T-ReCS imaging and spectroscopy. The image
reduction starts with sky subtraction, stacking of individual
images, and rejection of bad images. Next, flux calibration is performed 
using  the standard star. These were selected from the catalogue of 
spectrophotometric standard stars published by \cite{1999AJ....117.2428C}.
The final
step for image reduction is aperture photometry of the
target inside a default aperture radius of 0.9 arcsec,
which is a good estimate of the total flux for point-like sources.

\subsection{GTC/CC 7.5-13.5 $\mu$m spectroscopy}
\label{cc_spec}

\begin{table*}
	\begin{minipage}{1.\textwidth}
		\caption{Log of GTC/CC spectroscopic observations in $N$ band ($7.5-13.5$ $\mu$m). 
Column 1 gives the name of the QSO, column 2 the date of 
observations, column 3 the on-source time integration, column 4 the 
name of the spectrophotometric standard star, column 5 the FWHM of the 
standard star, columns 6 and 7 the airmass and the precipitable water vapor during the observations, column 8 the position angle  of the slit, and column 9 the observational programme for which the data were acquired.
}
	\label{spec_cc} 
		\centering
\begin{tabular}{lcccccccc} 
				\hline
				Name	& Date  &$t_{on}$ & STD     & $FWHM_{STD}$ &Airmass & PWV  & PA & Programme \\
				&  &   (s)      & & (arcsec)  & & (mm)   & (deg)   &  \\
				\hline
				\hline
				PG 0003+199 & 2013.09.22  & $2\times766$ & HD 2436  & 0.24  & 1.25 & NA & 65 & GT  \\        
				PG 0804+761  & 2014.03.15 	&  $3\times354$  & HD 64307 &  0.33 & 1.48 & 7.7 & 360& ESO-GTC \\
				& 2014.01.03 & $3\times354$ & HD 64307 & 0.33  & 1.56 & 7.6 & 360 &  \\             
				PG 0844+349  & 2014.12.03 & 1238  & HD 81146 & 0.28 & 1.45 & $4.5-4.1$ & 0  & ESO-GTC \\
				& 2014.12.05 & 1238 & HD 81146 & 0.34 & 1.15 & 7.2 & 360 &  \\
				PG 1211+143  & 2014.03.14 & $3\times295$  & HD 113996 & 0.34 & 1.07 & 8 & 0   & ESO-GTC\\
				& 2014.06.18 & 943 & HD 109511 & 0.24 & 1.31 & 8.1 & 360 &  \\
				PG 1229+204  & 2014.06.20 & 1238   & HD 111067  & 0.42 & 1.35 & 7.1 & 360 & ESO-GTC \\
				& 2014.06.20 & 1238 & HD 111067 & 0.42 & 1.31 & 8.1& 360 &  \\
				PG 1411+442  & 2014.05.30 & 1238 & HD 128902 & 1.05 &  6.9& 6.9 & 360  & ESO-GTC \\
				& 2014.05.31 & 1238 & HD 128902 & 0.36 & 1.35 &  5.9& 360 &  \\
				PG 1426+015  & 2014.05.19 & 943 & HD 126927 & 0.34 & 1.16 & $3.8-3.7$ & 360 &  ESO-GTC\\
				& 2014.06.09 & 943 & HD 126927 & 0.30 & 1.33 & 8.7       &       \\  
				PG 1440+356  & 2014.06.07 & 1238 & HD 128902 & 0.30 & 1.03& $5.8-6.9$& 360   & ESO-GTC  \\
				& 2014.06.07 & 1238 & HD 128902 & 0.26 & 1.12 & $6.1-6.5$& 360   &  \\ 
				PG 1501+106  & 2014.05.27 & 943  & HD 133165 & 0.37 & 1.12 & 4.8& 360   & ESO-GTC  \\
				& 2014.05.01 & 943  & HD 133165 & 0.28 & 1.33 & 6.0 & 360   &  \\			
				PG 2130+099      & 2014.09.21 & $2\times766$ & HD 206445 & 0.31 & 1.05&$4.2-5.6$  & 90 & GT \\ 
				MR 2251-178 & 2015.07.06 & 1120 & HD219449  & 0.52 & 1.52 & 7.0 & 90  &  MEX \\
				
				\hline
			\end{tabular}\\ 
		\end{minipage}
		
\end{table*}

We observed 10 QSOs within the guaranteed and ESO-GTC time in low-resolution 
spectroscopy ($R=175$) at $N$ band
($\lambda_{c}=7.5-13.5$ $\mu$m). These data were acquired between September
2013 and December 2014 with a slit width  of $0.52$~arcsec. 
Furthermore, Mrk~509 and PG~0050+124 (IZw1) were already observed with
VISIR/VLT by \citet{2010A&A...515A..23H} and \citet{2013A&A...558A.149B},
respectively (see Section \ref{ancillary}), and we used their 
high-resolution spectroscopy in our analysis. 
We also observed MR~2251-178 within the Mexican time because this is the 
only object in the sample without \emph{Spitzer/IRS}  spectra.  The low
resolution spectroscopy for this object was obtained in July 2015
with a slitwidth of $0.52$ arcsec.  Altogether, spectroscopy was acquired in 8-10m-class telescopes for 13 out of the 20 objects. 

On average, CC spectroscopy was obtained with a $PWV\sim6.6$ mm
and airmass $\sim1.27$. We estimate the 
image quality from the image of the standard star obtained at Si2 band 
just before the acquisition of the science target ($<FWHM>\sim0.4$~arcsec). The log of CC 
spectroscopic observations is shown in Table \ref{spec_cc}.

The spectroscopic observations were also reduced  with {\sc RedCan}.  
The first steps of the reduction process are similar to those for 
imaging, followed by two-dimensional wavelength
calibration of the target and standard star using sky lines.  Then,
we define the trace of a PSF using the observations of the
standard star. Finally, a point-like extraction was
made and we applied both slit-loss correction
and aperture
correction. For more details on the MIR data reduction pipeline 
 see \cite{2013A&A...553A..35G}.

\section{Ancillary data} 
\label{ancillary}

We compile high angular resolution NIR ($1-3$ $\mu$m), MIR ($5-35$
$\mu$m) imaging and spectroscopy from the literature with estimates of
unresolved emission, when available, in order to 
build complete nuclear NIR-to-MIR SEDs of the
QSOs in the sample. The unresolved fluxes are the
result of removing, using various methods, the underlying NIR and MIR
emission related to the host galaxy, and not directly linked to AGN and 
dusty torus emission. 
In Table \ref{NIR_data} we list the data gathered from previous studies.

\begin{table*}
		\caption{NIR unresolved fluxes from the literature. Column 1 gives the names of the QSOs, column 2 the unresolved flux at $H$ band from HST/NICMOS \citep{2006ASPC..357..231V, 2009ApJ...701..587V}, and columns 3 and 4 the 
upper limits at $K$ and $H$ bands \citep{2001AJ....122.2791S}, respectively.\label{NIR_data}}
		\begin{minipage}{1.\textwidth}
		\begin{center}
			\begin{tabular}{l|cccc} 
					\hline\hline \\
					Name  	& F160W-HST/NICMOS (PSF)  & K$'$-QUIRC/Gemini (PSF) & H-QUIRC/Gemini (PSF) \\
					& $f_{\nu}$ (mJy)     &   $f_{\nu}$ (mJy) &  $f_{\nu}$ (mJy)     \\
					\hline
					\hline
					PG 0007+106 & 2.9   & $<19$.3 & ...\\
					PG 0050+124 & 11.4   & $<3$9.6 & ...    \\
					PG 0804+761 & ... & $<26$.2 &$<13$.2\\
					PG 0844+349 & 5.9  & ...&...\\
					PG 1211+143 &  ...     & ... & $<11$.51 \\
					PG 1229+204 & 2.7  & $<7$.0 &... \\
					PG 1351+640 & ...    & $<9$.3 & $<6.$4 \\
					PG 1411+442$^{a}$ & 7.1 & $<15$.8 &... \\
					PG 1426+015 & 5.8  & $<18$.1 & ...\\
					PG 1440+356 &  9.1 & $<17$.6 & ...\\
					PG 2130+099 & 9.2 & $<25$.5 & ...\\
					PG 2214+139 & 6.5 &  ...   &  ...\\
					Mrk 509$^{b}$  & ... & ... &  ...\\
					\hline
				\end{tabular} \\
				\end{center}
			{\bf Notes}.--$^{a}$This object presents an unresolved flux at $J$ band $<6.44$ mJy from QUIRC/Gemini . 
$^{b}$This object presents fluxes measured in an aperture diameter of 3 arcsec at $J$ ($<10.7$ mJy), 
$H$ ($<14.1$ mJy) and $K$ ($<22.3$ mJy) bands from ISAAC/VLT, which have been used in the present work as upper limits \citep{2006A&A...452..827F}. 

			\end{minipage}

\end{table*}

\subsection{HST/NICMOS data}

 \cite{2006ASPC..357..231V} and \cite{2009ApJ...701..587V} observed with NICMOS/\emph{HST} at $H$ band 
(F160W, $\lambda_{c}=1.60$ $\mu$m) a sample of 28 QSOs as part of the Quasar/ULIRG Evolution 
Study (QUEST), 9 of which are in our sample. The high angular resolution ($\sim0.3$ arcsec) and 
pixel size ($\sim0.076$ arcsec pixel$^{-1}$) of NICMOS allows to 
have a good estimate of the unresolved 
emission ($FWHM=0.14$ arcsec). In the present work we have used the unresolved emission reported in this band. The photometric errors are $\sim10\%$.

\subsection{Upper limits from QUIRQ on Gemini North and ISAAC/VLT}

\cite{2006ApJS..166...89G} report unresolved emission at $K'$ band  ($\lambda_{c}=2.12$ $\mu$m,
$\Delta_{\lambda}=0.41$ $\mu$m), and at $H$ band 
($\lambda_{c}=1.65$, $\Delta_{\lambda}=0.30$ $\mu$m) for several of our QSOs (see Table \ref{NIR_data}). These data were obtained with the IR camera
QUIRC (Hodapp et al. 1996) on the Gemini North telescope, with an
angular resolution $\sim0.2$~arcsec. 

Mrk~509 has nuclear fluxes at $Js$ ($\lambda_{c}=1.25$ $\mu$m), $H$
($\lambda_{c}=1.65$ $\mu$m) and $Ks$ ($\lambda_{c}=2.16$ $\mu$m) bands
measured with the IR camera ISAAC on the Very Large Telescope (VLT), with a
spatial resolution of $0.6-1$ arcsec. The nuclear flux was, however, measured in an aperture diameter of 3 arcsec
\citep{2006A&A...452..827F}, and hence, we use these fluxes as upper limits.

\subsection{VISIR/VLT nuclear spectroscopy at N band}

PG~0050+124 and Mrk~509 have $N$-band ($\lambda_{c}=10$ $\mu$m) 
low-resolution ($R=300$) spectroscopy acquired with
VISIR, with a spatial resolution $\sim0.3$ arcsec
\citet{2013A&A...558A.149B, 2010A&A...515A..23H}. The slitwidth was 0.75
arcsec and the spectra cover a wavelength s between $\sim7.5-13.5$
$\mu$m.

\subsection{\emph{Spitzer/IRS} spectroscopy}

Most QSOs in our sample are part of the 
\emph{Spitzer/IRS} telescope spectroscopic survey 
QUEST (PID: 3187; PI: Veilleux). In
general, they have been observed with  the
following low-resolution modes\footnote{{\url http://irsa.ipac.caltech.edu/}}: SL1$\sim7.4-14.5$ $\mu$m,
SL2$\sim5.2-7.7$ $\mu$m, LL1$\sim19.9-39.9$ $\mu$m and
LL2$\sim13.9-21.3$ $\mu$m. The slit widths range from 3.6 to 11.1~arcsec. 

Fully-reduced and calibrated 
spectra were downloaded from the  Cornell Atlas database of \emph{Spitzer/IRS}  CASSIS \citep[v6;][]{2011ApJS..196....8L}, which provides
 optimal extraction regions to ensure the best $S/N$
ratio. We stitch the different module spectra together using 
module SL2 as a reference spectrum for flux scaling using our own {\sc Python} routines. 

\section{Analysis}
\label{analysis}
\subsection{MIR imaging photometry at Si2 band ($\lambda_{c}=8.7 \,\mu$m)}
\label{cc_image}

\begin{figure*}
\begin{minipage}{1.\textwidth}
\begin{tabular}{cc}
\centering
\includegraphics[scale=0.6]{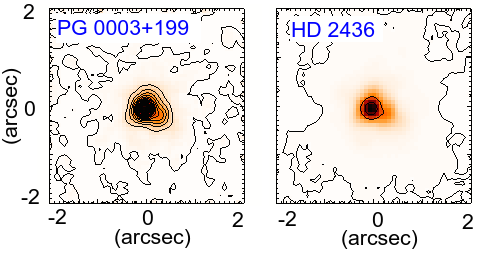}&\includegraphics[scale=0.2]{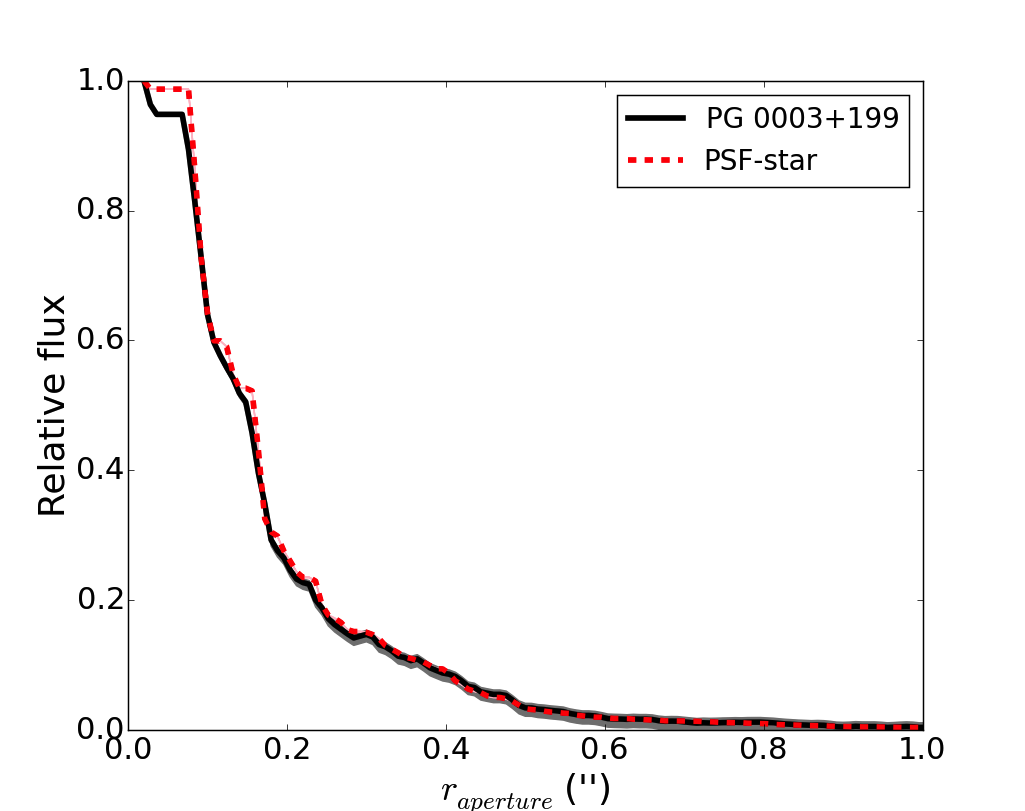}
\end{tabular}
\caption[kk2]{Left and middle panels: Si2 images of PG~003+199 and its standard star. The lowest contour is $3\sigma$ over the background and the next contours are traced in $2\sigma$ steps (except in the PSF image). Right panel: radial profiles of PG~0003+199  
in black solid line, 
and its standard star in red dotted line.
 \label{cc_2}
} 
\end{minipage}
\end{figure*}

\begin{figure*}
\begin{minipage}{1.\textwidth}
\begin{tabular}{cc}
\centering
\includegraphics[scale=0.5]{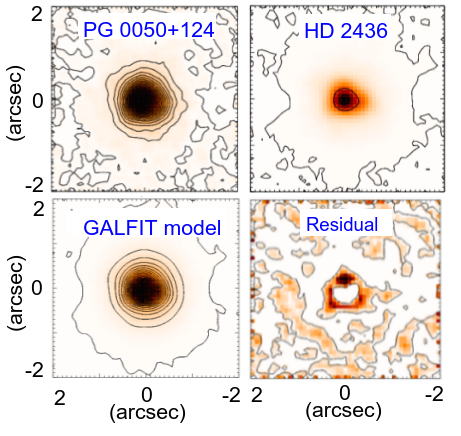}&\includegraphics[scale=0.3]{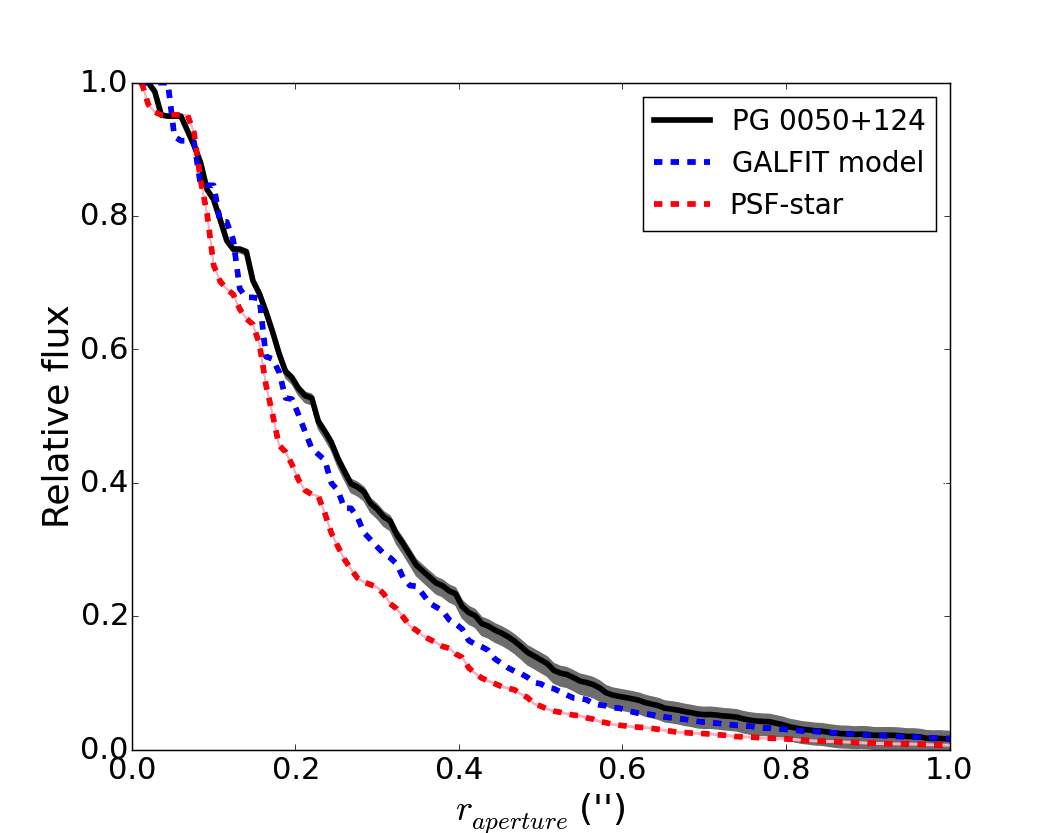}
\end{tabular}
\caption[kk2]{Left to middle panels: Si2 images of  PG~0050+124 and 
its standard star, {\sc GALFIT} model and its residual image. 
Right panel: radial profile of PG~0050+124 in black solid line, 
its standard star in red dotted line, and best {\sc GALFIT} model in 
blue dotted line.
\label{cc_1}
} 
\end{minipage}
\end{figure*}

In order to estimate if the QSOs have extended emission over the
stellar PSF, we perform aperture photometry on the QSOs and their
corresponding standard stars with increasing apertures using  
the  {\sc phot} and {\sc aphot} tasks of  the image
analysis package {\sc IRAF}. We combine these measurements  to
build radial profiles. Figure \ref{cc_2} shows the 
images of PG~0003+199, its standard star and their radial profiles, in 
which it is clearly seen that the QSO is dominated by 
unresolved emission. In Figure \ref{cc_1} we show the data and analysis for 
PG~0050+124, which presents a
clear extended component. For this object we also show the {\sc
  galfit} model and its residual (see Appendix \ref{individual} for
more details on the analysis of the extended component with {\sc
  galfit}).  The images and analysis for the rest of the QSOs  can be seen in 
Appendix \footnote{\url{https://drive.google.com/open?id=0B9f3R8mc1H8BN3Y0b3RIWDViZ2M}}.

The uncertainties in the radial profiles include 
photometric errors, which are estimated as $\sqrt{\sigma_{\rm
    back}^{2}N_{\rm pix} + \sigma_{\rm
    back}^{2}N_{\rm pix}^{2}/N_{\rm pix-ring}}$, where
$N_{\rm pix}$ is the number of pixels inside the aperture considered, $N_{\rm pix-ring}$ is the number of 
pixels inside an 80 pixel-width ring around the source, 
used to estimate the background level and its
standard deviation $\sigma_{\rm back}$ \citep[see][]{2005PASP..117..978R}. The second term of the error equation is 
almost negligible because the backgrounds in our 
images are flat. In addition, we also consider a six per cent
uncertainty due to time-variability of the sky transparency and adopt a 13~per cent uncertainty due 
to PSF-variability\footnote{This has been calculated from the analysis 
of several standard stars observed during the same night at MIR with T-ReCS on Gemini \citep{2012AJ....144...11M}.}.  

In order to estimate variations of the PSF due to variable sky
conditions, we use observations of the same standard star acquired
consecutively.  For two of the QSOs (PG~1426+015 and PG~1229+204) we
acquired several observations of the standard star just before or
after the target.  We find that the FWHM of the PSF varies $\sim15\%$ in time scales of a few minutes. Consistent with the results of \citet{2012AJ....144...11M}. This is likely the reason
why several QSOs (e.g. PG~0804+761, PG~0844+349, PG~0923+129,
PG~1229+204, PG~1351+640, PG~2130+099) show radial profiles with FWHM slightly
narrower than their corresponding standard stars (see Figures in
Appendix$^{{\color{blue}4}}$).

Allowing for these mild FWHM variations, 
the radial profile of the standard stars 
represents the maximum
contribution to the emission due to an unresolved source. Therefore,
taking into account all the uncertainties, all QSOs, except PG~0050+124, are
unresolved in our MIR imaging (see Figures in Appendix$^{{\color{blue}4}}$).
In Table \ref{Si2_flujos} we list the flux at Si2 band 
measured inside an aperture radius of 1 arcsec, plus the ratio between
the FWHM of the QSO and its standard star, which shows that the majority
of the objects in the sample are indeed unresolved. The errors are estimated 
adding in quadrature the
photometric error, the flux calibration, the time-variability and the
PSF-variability uncertainties.

	\begin{table} 
			\caption{Results of the imaging analysis. Column 1, name of the QSO; column 2, unresolved flux at Si2 (8.$7\, \mu$m) band; 
column 3, ratio between the QSO and standard star FWHM.\label{Si2_flujos} } 
			\centering
			\begin{tabular}{lccc} 
				\hline\hline 
				Name	& $f_{\nu}$ ($r=1''$)  & $\frac{FWHM_{target}}{FWHM_{STD}}$ \\
				& (mJy)     &        \\
				\hline
				\hline
				PG 0003+199 &   159.$2\pm0$.1 & 1.0 \\
				PG 0007+106 &   40.$8\pm0$.2  & 1.2 \\
				PG 0050+124 &   249.$1\pm0$.2 & 1.3 \\
				PG 0804+761 & 105.$1\pm0$.2 & 1.0  \\
				PG 0844+349 &  29.$4\pm0$.7 & 0.9 \\
				PG 0923+129 &  92.$6\pm0$.2 & 0.9  \\
				PG 1211+143 &   89.$3\pm0$.2 & 1.1 \\
				PG 1229+204 &   31.$5\pm0$.2 & 1.0  \\
				PG 1351+640 &    67.$2\pm0$.2 & 0.9  \\
				PG 1411+442 &   65.$7\pm0$.4 & 1.0  \\
				PG 1426+015 &   56.$0\pm0$.1 & 1.1   \\
				PG 1440+356 &   67.$200\pm0$.004  & 1.3 \\
				PG 1448+273 &   25.$0\pm0$.2  & 1.1   \\
				PG 1501+106 &   92.$2\pm0$.1 & 1.0 \\
				PG 1534+580 &  111.$9\pm0$.9 & 1.2 \\
				PG 1535+547 &   43.$7\pm1$.0 & 1.1  \\
				PG 2130+099 &  118.$7\pm0$.1 & 0.9  \\ 
				MR2251-178  &   49.$2\pm0$.2 & 1.3  \\
				\hline
			\end{tabular} 
	\end{table}

\subsection{\emph{Spitzer/IRS}  and GTC/CC spectroscopy}
\label{spec}

We compare the nuclear GTC/CC and \emph{Spitzer/IRS}  spectra for the 10 QSOs
with both types of data (namely, PG~0003+199, PG~0804+761,
PG~0844+349, PG~1211+143, PG~1229+204, PG~1411+442, PG~1426+015,
PG~1440+356, PG~1501+106 and PG~2130+099, see Figure \ref{Sed1} and Figures in Appendix$^{{\color{blue}4}}$). We observe that, within
the uncertainties, the shapes of both spectra are similar. 
The most notable differences are at the edges
of the GTC/CC spectra ($\sim7.5$ and $\sim13.5\,\mu$m)
and in the 
$\sim9.0-9.7$ $\mu$m rest-frame range. We  attribute these differences 
to low atmospheric
transmission\footnote{\url{http://www.gtc.iac.es/instruments/canaricam/MIR.php}}
(see Figure \ref{Sed1}).

In general, the flux measured in the Si2 band within 1~arcsec apertures 
is consistent with both the nuclear GTC/CC and \emph{Spitzer/IRS}  spectra.  For MR~2251-178, the only QSO that does not have \emph{Spitzer/IRS} spectrum, the unresolved emission in Si2 band is consistent with the
spectral flux at 8.7 $\mu$m. 
As an example, in
Figure \ref{Sed1} we show the observed NIR to MIR unresolved SED and
spectroscopy of PG~1411+442, and the rest 
of the QSOs are show in Appendix$^{{\color{blue}4}}$.

\begin{figure}
\hspace{-1cm}
\includegraphics[scale=0.3]{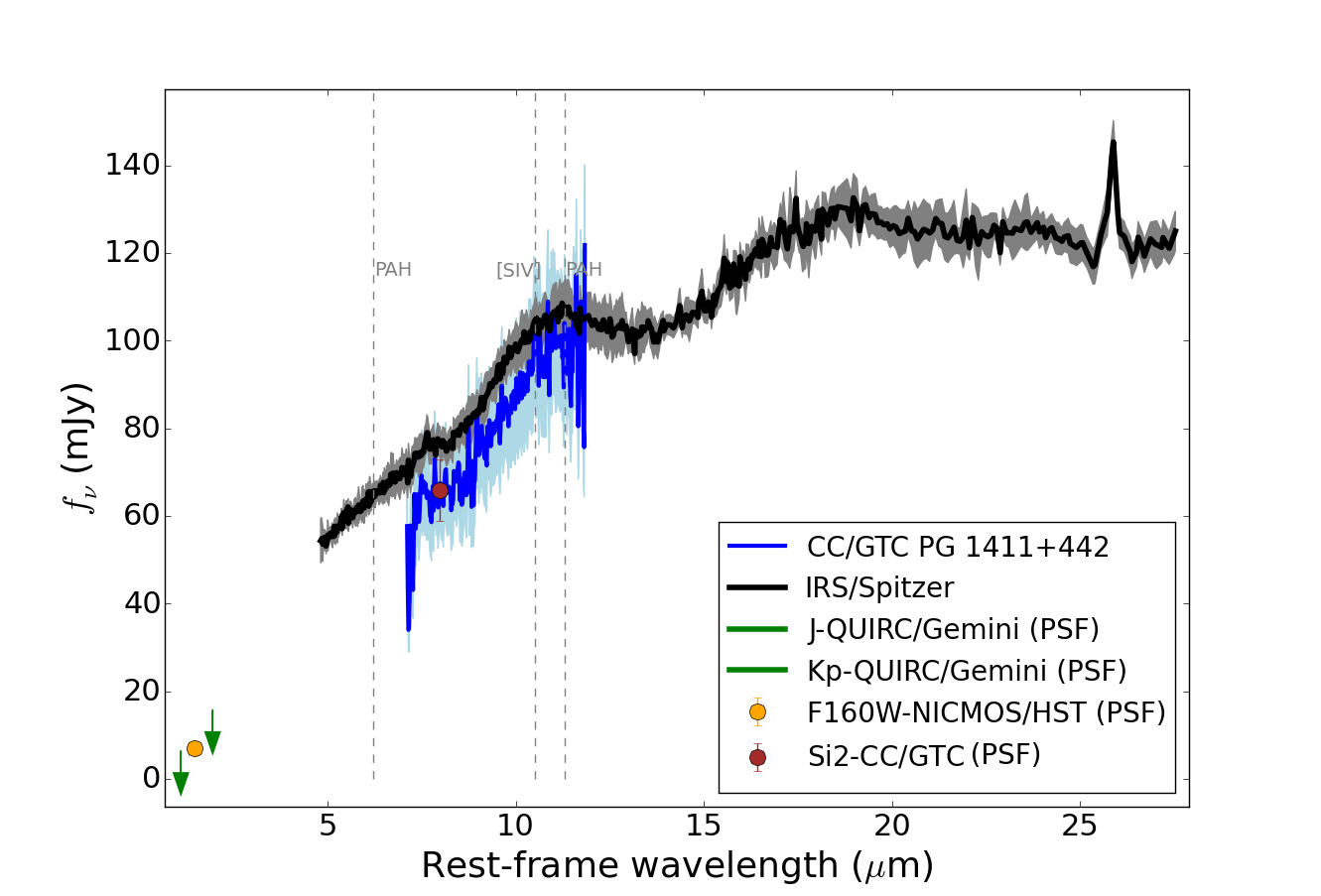}	
\caption{SED of the unresolved component of PG~1411+442, using our
GTC/CC unresolved emission, the ancillary photometry, the GTC/CC spectrum 
(nuclear, width slit $\sim0.5''$) 
and the \emph{Spitzer/IRS} spectrum (width slit $\sim3''$). 
\label{Sed1}}
\end{figure}

For the QSOs that do not have nuclear GTC/CC
spectra we compare the flux of the unresolved emission in Si2-band image with that derived from the
\emph{Spitzer/IRS} spectra. PG~1351+640 and PG~1535+547 have Si2
unresolved emission consistent with their \emph{Spitzer/IRS} spectra at
8.7 $\mu$m (see Figures in Appendix$^{{\color{blue}4}}$). PG~0007+106 and PG~1448+273 have unresolved emission
fainter than their spectra at 8.7 $\mu$m, suggesting the presence of an
extended component within the {\it Spitzer} aperture 
(see Figures in Appendix$^{{\color{blue}4}}$).
Finally, PG~0923+129 and PG~1534+580 have unresolved emission estimated
from the image that is brighter than that of its spectrum at 8.7 $\mu$m (see Figures in Appendix$^{{\color{blue}4}}$), probably due to uncertainties in the flux loss correction \citep{2016MNRAS.455..563A}.
PG~2214+139 has a bad quality image at Si2 band and it is not possible to measure 
the flux to compare it with its \emph{Spitzer/IRS} spectrum. 

For PG~0050+124 and Mrk~509 we compare the nuclear VISIR/VLT spectra
with the \emph{Spitzer/IRS} spectra and the unresolved emission at Si2
band from CC. In both cases the flux of the nuclear VISIR/VLT spectra
is lower than the \emph{Spitzer/IRS} ones, while the CC unresolved emission
is consistent with the nuclear VISIR/VLT spectra (see Figures in  Appendix$^{{\color{blue}4}}$. The spectral shapes of
VISIR/VLT and \emph{Spitzer/IRS}  spectra are also similar.

The nuclear and \emph{Spitzer/IRS}  spectral shapes are
similar and the unresolved emission at Si2 band is consistent
with the nuclear and \emph{Spitzer/IRS}  spectra. Thus, we assume that the
\emph{Spitzer/IRS}  spectra of all QSOs in our sample are mostly
dominated by emission due to the AGN and its surrounding torus.

\begin{table*} 
		\caption{ Spectral decomposition.
 Column 1 gives the name of the QSO, columns 2 and 3 the name of the 
PAH template and its contribution, 
columns 4 and 5 the name of the AGN template and its contribution between 5 and 15 $\mu$m within the IRS slit, columns 6 and 7 the name of the stellar template and its contribution, and columns 8 and 9 the reduced $\chi_{\nu}^{2}$ and the rms error coefficient for the best-fitting model. The fractional contribution is measured between $5-15$ $\mu$m.}\label{Spec_decomp1} 
		\scalebox{0.9}{
			\begin{tabular}{l|cccccllc} 
				\hline\hline 
				
				Name	& PAH  	& \%	& AGN 	&	\%	&	stellar emission	&	\%	&	$\chi_{\nu}^{2}$	&	$CV_{RMSE}$	\\

				\hline
				PG 0003+199	&	NGC 3769	&	0.3	&	PG 1114+445	&	83.6	&	M 85	&	16.0	&	0.071	&	0.019	\\
				PG 0007+106	&	NGC 3187	&	3.1	&	PG 1114+445	&	88.4	&	NGC 5812	&	8.5	&	0.226	&	0.030	\\
				PG 0050+124& NGC 2993	& 0.0 	& [HB89] 1402+436	&	99.5 	& M 85	&	0.5 	&	0.45 	&	0.042 	\\
				PG 0804+761	&	NGC 3769	&	0.1	&	J143220.15+331512.2	&	75.2	&	NGC 1700	&	24.7	&	0.315	&	0.042	\\
				PG 0844+349	&	NGC 5996	&	3.3	&	J131217.7+351521	&	89.3	&	NGC 1374	&	7.5	&	0.289	&	0.036	\\
				PG 0923+129	&	UGC 09618	&	7.0	&	PG 1149-110	&	88.8	&	M 85	&	4.2	&	0.279	&	0.031	\\
				PG 1211+143 & NGC 3769	& 1.4 	& 3C 445	&	98.6	& NGC 4474	&	0.0 	&	1.043 	&	0.023 	\\
				PG 1229+204	&	NGC 2993	&	0.0	&	PG 1149-110	&	87.5	&	NGC 1700	&	12.5	&	0.227	&	0.031	\\
				PG 1351+640 & NGC 2993	& 0.0 	& 2MASX J02343065+2438353	&	97.1 	& NGC 1700	&	2.9 	&	4.33 	&	0.146 	\\
				PG 1411+442	&	J14361112+6111265	&	1.7	&	J1640100+410522	&	80.2	&	NGC 1700	&	18.1	&	0.096	&	0.019	\\
				PG 1426+015	&	ESO 557-G-001	&	1.0	&	VII Zw 244	&	86.0	&	NGC 1700	&	13.0	&	0.120	&	0.027	\\
				PG 1440+356	&	ESO 467 G 013	&	8.1	&	Mrk 1501	&	86.3	&	NGC 0821	&	5.6	&	0.530	&	0.021	\\
				PG 1448+273	&	NGC 3310	&	3.4	&	J160222.38+164353.7	&	83.7	&	NGC 5812	&	12.9	&	0.333	&	0.048	\\
				PG 1501+106 & MCG +08-11-002 	& 0.0 	& PG 1149-110	&	99.2 	& NGC 5831	&	0.7 	&	0.2 	&	0.029 	\\
				PG 1534+580	&	ESO 244-G-012	&	2.5	&	VII Zw 244	&	93.8	&	NGC 5831	&	3.7	&	0.271	&	0.032	\\
				PG 1535+547	&	NGC 3769	&	2.4	&	J1640100+410522	&	80.1	&	NGC 1700	&	17.5	&	0.203	&	0.028	\\
				PG 2130+099	&	NGC 3187	&	0.6	&	J14492067+4221013	&	89.1	&	NGC 4570	&	10.3	&	0.107	&	0.025	\\
				PG 2214+139	&	NGC 2993	&	0.0	&	HB89-1435-067	&	84.8	&	NGC 4570	&	15.2	&	0.062	&	0.016	\\
				Mrk 509	&	NGC 3310	&	8.1	&	3C 390.3	&	84.0	&	NGC 1549	&	7.9	&	0.068	&	0.023	\\
				\hline\\
			\end{tabular}
}  		
\end{table*}

\begin{table*}
	\begin{minipage}{.6\textwidth}
		\caption{ Parameters of the AGN component obtained for the \emph{Spitzer/IRS} spectral decomposition. Column 1: name of the QSO. Column 2: AGN luminosity at 6 $\mu$m. Column 3: AGN luminosity at 12 $\mu$m. Column 4: MIR spectral index ($8.1-12.5$ $\mu$m). Column 5:  silicate strength index. \label{Spec_decomp2} } 
\begin{tabular}{l|cccccccc} 
				\hline\hline 
				Name	& $log L_{6 \, \mu m}=L_{6}$ & $log L_{12 \, \mu m}=L_{12}$  &  $\alpha$ (8.$1-12$.5 $\mu$m)& $S_{sil}$ \\
				& (erg s$^{-1}$)	& (erg s$^{-1}$)   &		&					 \\			
				\hline
				PG 0003+199	&	$43.70_{-0.5}^{+0.15}$	&		$43.77_{-0.08}^{+0.06}$	&		$-1.5^{+0.6}_{-0.8}$		&	$0.1_{-0.3}^{+0.4}$	 	\\
				PG 0007+106	&	$44.38_{-0.24}^{+0.14}$	&		$44.46_{-0.08}^{+0.06}$	&		$-1.5_{-0.7}^{+0.5}$		&	$0.01_{-0.30}^{+0.30}$		\\
				PG 0050+124$^{a}$ & $44.82_{-0.03}^{+0.03}$     &  $44.95_{-0.03}^{+0.02}$         &  $-1.4_{-0.3}^{+0.1}$   & $0.2_{-0.2}^{+0.2}$  \\
				PG 0804+761	&	$44.88_{-0.14}^{+0.10}$	&		$44.89_{-0.05}^{+0.04}$	&		$-1.1_{-0.4}^{+0.4}$		&	$0.3_{-0.3}^{+0.2}$		\\
				PG 0844+349	&	$43.92_{-0.23}^{+0.13}$	&		$44.04_{-0.07}^{+0.05}$	&		$-1.6_{-0.6}^{+0.5}$		&	$0.2_{-0.3}^{+0.2}$		\\
				PG 0923+129	&	$43.42_{-0.23}^{+0.14}$	&		$43.69_{-0.07}^{+0.05}$	&		$-1.9_{-0.7}^{+0.5}$		&	$-0.03_{-0.30}^{+0.20}$		\\
				PG 1211+143   & $44.60_{-0.22}^{+0.12}$           &  $44.91_{-0.01}^{+0.01}$             &     $-1.9_{-0.8}^{+0.5}$              &  $-0.04_{-0.40}^{+0.20}$   \\
				PG 1229+204	&	$43.90_{-0.26}^{+0.15}$	&		$44.10_{-0.06}^{+0.05}$	&		$-1.9_{-0.8}^{+0.5}$		&	$-0.04_{-0.40}^{+0.20}$	 	\\
				PG 1351+640$^{a}$ & $44.64_{-0.1}^{+0.01}$     &  $44.72_{-0.07}^{+0.06}$ &  $-2.0_{-0.1}^{+0.1}$                & $0.6_{-0.1}^{+0.1}$   \\
				PG 1411+442	&	$44.76_{-0.33}^{+0.5}$	&		$44.47_{-0.11}^{+0.09}$	&		$-1.4_{-0.7}^{+0.6}$		&	$0.0_{-0.4}^{+0.3}$		\\
				PG 1426+015	&	$44.49_{-0.21}^{+0.13}$	&		$44.64_{-0.07}^{+0.05}$		&	$-1.7_{-0.7}^{+0.6}$		&	$0.07_{-0.30}^{+0.30}$	 	\\
				PG 1440+356 &	$44.33_{-0.32}^{+0.18}$	&		$44.38_{-0.13}^{+0.09}$		&	$-1.6_{-0.9}^{+0.7}$		&	$-0.03_{-0.40}^{+0.30}$		\\
				PG 1448+273	&	$43.84_{-0.62}^{+0.13}$	&		$44.08_{-0.07}^{+0.05}$		&	$-1.8_{-0.7}^{+0.5}$		&	$0.0_{-0.3}^{+0.3}$		\\
				PG 1501+106 & $43.80_{-0.62}^{+0.12}$     &  $44.10_{-0.04}^{+0.04}$ &   $-2.1_{-0.6}^{+0.5}$  & $-0.1_{-0.4}^{+0.3}$      \\
				PG 1534+580	&	$43.34_{-0.25}^{+0.14}$		&	$43.60_{-0.06}^{+0.05}$	&		$-1.9_{-0.7}^{+0.4}$		&	$0.03_{-0.30}^{+0.30}$		\\
				PG 1535+547	&	$43.58_{-0.27}^{+0.16}$		&	$43.59_{-0.09}^{+0.08}$	&		$-1.4_{-0.7}^{+0.6}$		&	$0.07_{-0.30}^{+0.30}$	 	\\
				PG 2130+099	&	$44.38_{-0.26}^{+0.16}$		&	$44.50_{-0.10}^{+0.07}$	&		$-1.6_{-0.8}^{+0.6}$		&	$-0.07_{-0.40}^{+0.30}$	 	\\
				PG 2214+139	&	$44.17_{-0.27}^{+0.19}$		&	$44.17_{-0.11}^{+0.09}$	&		$-1.3_{-0.8}^{+0.6}$		&	$0.2_{-0.4}^{+0.3}$		\\
				Mrk 509   	&	$44.04_{-0.30}^{+0.17}$	&		$44.18_{-0.09}^{+0.07}$	&		$-1.8_{-0.8}^{+0.6}$		&	$-0.03_{-0.40}^{+0.30}$	 	\\    
				
				\hline
			\end{tabular}\\
			{\bf Notes}.-$^{a}$These objects were poorly fitted around the silicate feature, and hence we measure the feature on the spectra directly 
(see statistics in Table \ref{Spec_decomp1}).
		\end{minipage}				
\end{table*}

\section{\emph{Spitzer/IRS} spectral decomposition: isolating the AGN emission}
\label{results1}
Evidence for star formation through the detection of PAH features in the \emph{Spitzer/IRS} spectra is present in 40 per cent of the QSOs studied by \cite{2006ApJ...649...79S}, and also in the stacked spectrum of those that lacked individual detections, implying that starbursts are present in most QSOs. We use the tool {\sc deblendIRS}
\citep{2015ApJ...803..109H} to decompose the \emph{Spitzer/IRS}
spectra of the QSOs into their starburst (PAH and stellar) and AGN
components.  {\sc deblendIRS} uses a set of \emph{Spitzer/IRS}
templates of galaxies dominated by AGN emission, PAH emission 
(interstellar medium, ISM) and stellar emission (passive population of
the host galaxy) and  follows a $\chi^{2}$ minimization method to find
the combination of templates that best reproduces the spectrum of the 
source under study. Additionally, the rms-variation coefficient is also used as a 
second criterium to select the best combination of templates \citep{2015ApJ...803..109H}.

Using Bayesian inference, {\sc deblendIRS} estimates 
the probability distribution of the fractional contribution to the 
integrated MIR emission for the stellar, 
PAH and AGN components, the AGN luminosity contribution at 12 and 6 $\mu$m, 
and the starburst luminosity contribution at 12 $\mu$m. 
It also calculates the MIR spectral index $\alpha$ and the 
silicate strength $S_{sil}$ for the AGN component, where 
$\alpha$ is calculated 
between 6 and $12\,\mu$m assuming a power law 
$f_{\nu}=\nu^{\alpha}$, and the silicate strength is defined as 
\begin{equation}
	S_{sil}=\ln\frac{F(\lambda_{p})}{F_{C}(\lambda_{p)}},
\end{equation}
where $F(\lambda_{p})$ and $F_{C}(\lambda_{p})$ are the flux densities 
at the peak of the silicate feature and its underlying continuum, respectively. See \citet[][]{2015ApJ...803..109H} for details.

	\begin{figure}
	\hspace{-1cm}	\includegraphics[scale=0.45]{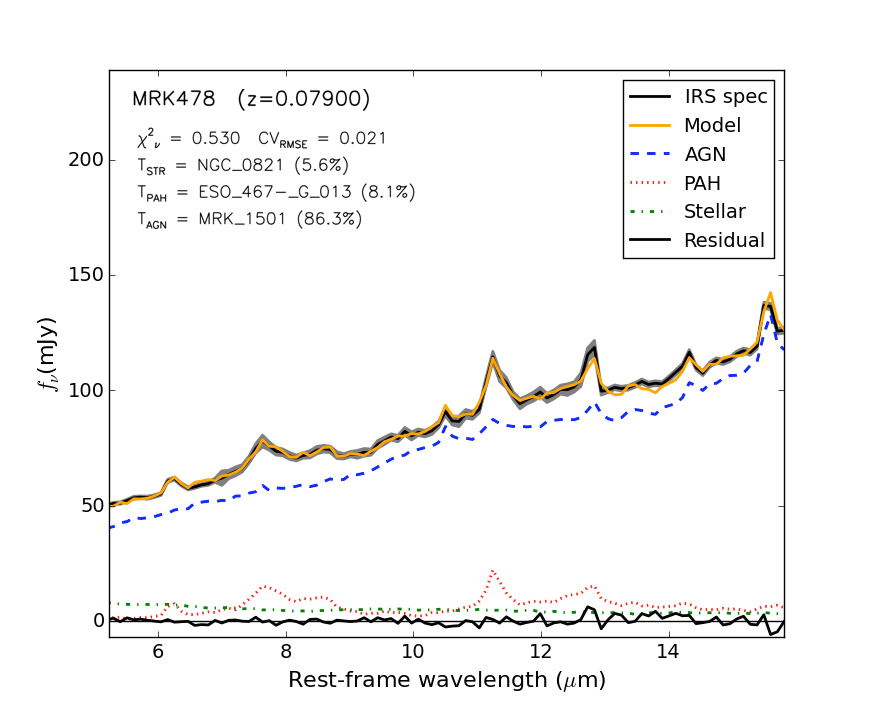} \\
		\includegraphics[scale=0.1]{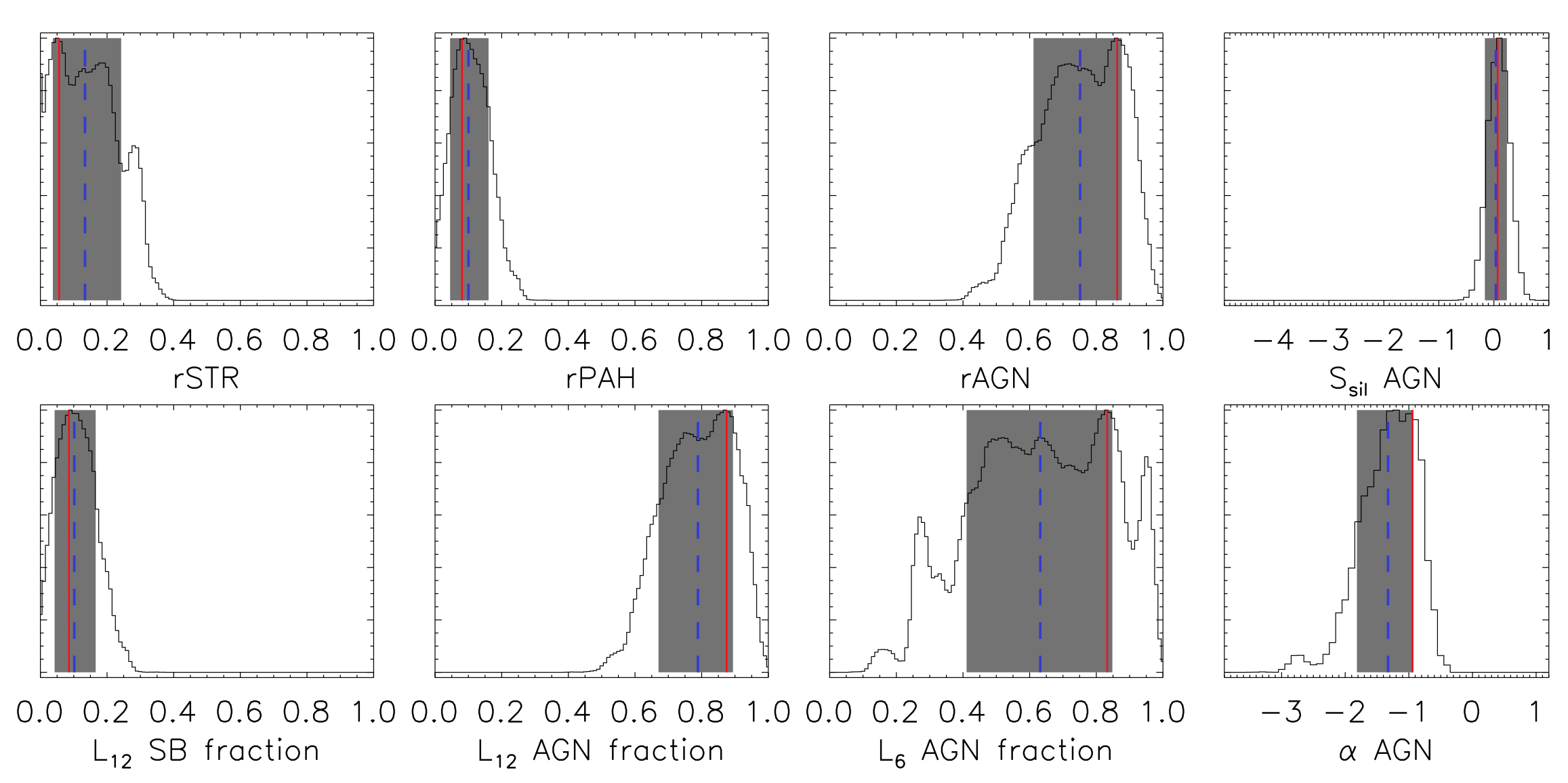}
		\caption{Output of the {\sc deblendIRS} spectral decomposition of PG 1440+356.  {\bf
                    Upper panel}: The black line shows the \emph{Spitzer/IRS} spectrum and the grey area its $1\sigma$
                  uncertainty. The orange line shows the best fitted
                  model, which is the sum of the AGN (blue
                  dashed line), PAH (red dashed line) and stellar
                  templates (dashed green line).  We also show the
                  residual of the fit as a black solid line around zero flux
                  density. {\bf Lower panels}: probability
                  distributions of the eight parameters obtained from
                  the spectral decomposition,  rSTR, rPAH and rAGN
                  stand for fractional contributions of the stellar,
                  PAH and AGN components, respectively; $S_{Sil}$~AGN
                  for silicate strength; $L_{12}$ SB fraction,
                  $L_{12}$ AGN fraction, $L_{6}$ AGN fraction for the
                  starburst and AGN fractional luminosities at 12 and
                  6~$\mu$m; and $\alpha$~AGN MIR spectral index of the AGN
                  component.\label{decomp_pg1440}}
	\end{figure}

	\begin{figure}
	\hspace{-1cm}	\includegraphics[scale=0.4]{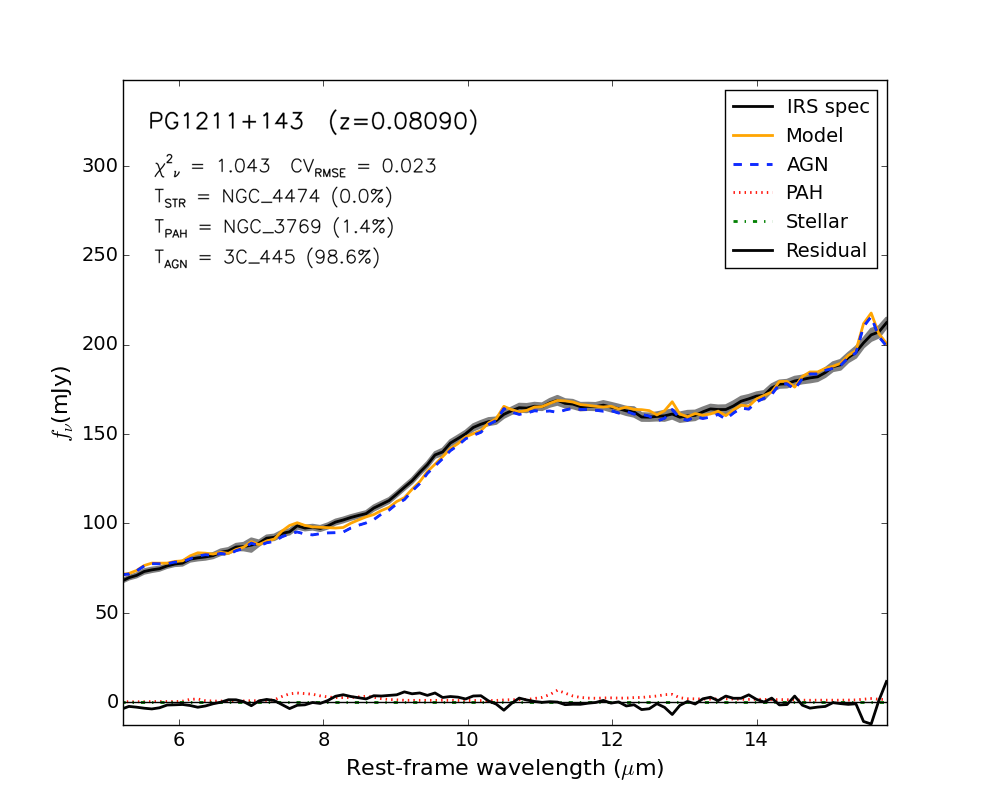}\\
		\includegraphics[scale=0.1]{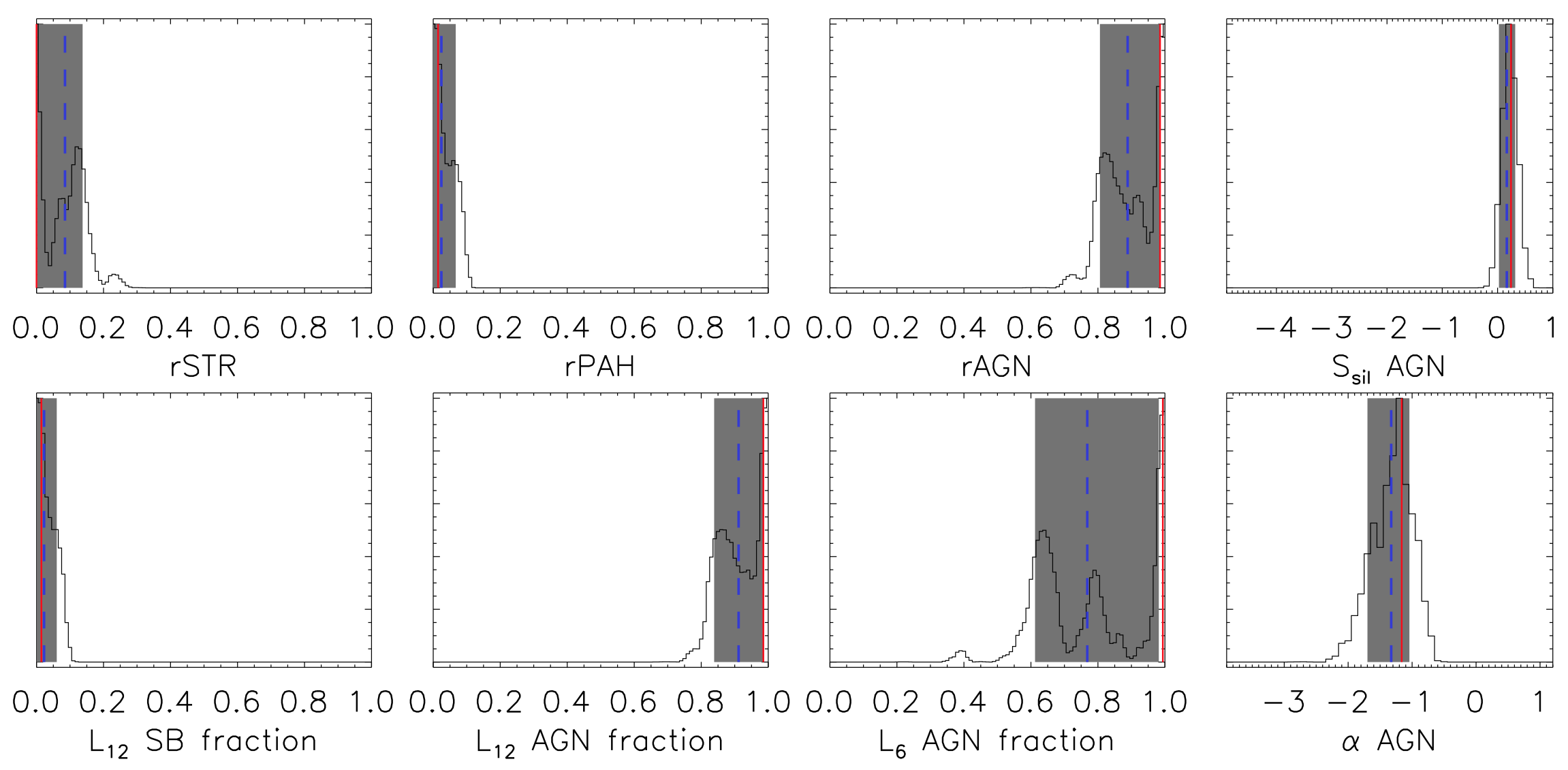} 
		\caption{  As Figure \ref{decomp_pg1440} but for PG 1211+143. Note that in this case the \emph{Spitzer/IRS is completely dominate by the AGN component (blue line), while the stellar and PAH components are negligible (green and red lines, respectively).}\label{decomp_pg1211}}
	\end{figure}

The spectral decomposition is done between $\sim5$ and 15 $\mu$m in the 
rest-frame.  
All 
spectra were re-sampled to a common wavelength resolution $\Delta\lambda=0.1$ $\mu$m.  In Figures \ref{decomp_pg1440} and \ref{decomp_pg1211} we show the 
spectral decomposition of a QSO dominated by the AGN component (Mrk~478)
and one with significant contributions by PAH and stellar components (PG~1211+143). Table~\ref{Spec_decomp1} lists the
results of the median and 68~per cent confidence intervals of the 
integrated MIR luminosity at $5-15$ $\mu$m, and within the \emph{Spitzer/IRS} slit, attributed to the three 
components, and the best templates used to decompose each QSO.
Table~\ref{Spec_decomp2} lists the 
AGN luminosity contribution at 12 and 6 $\mu$m, 
the starburst luminosity contribution at 12 $\mu$m, 
the MIR spectral index $\alpha$ and the 
silicate strength $S_{sil}$. PG~0050+124 and PG~1351+640 were poorly fitted around the silicate feature, and hence we measure the feature on the spectra directly.

We cannot find a combination of templates to
reproduce the spectrum of PG~1351+640
(see Figure in Appendix$^{{\color{blue}4}}$), 
probably because this object presents prominent silicate feature in emission. 
This object, together with PG~0050+124,
PG~1211+143 and PG~0804+761, are among the first  objects in
which prominent emission of silicates at 10 and 18 $\mu$m was observed
\cite[eg.,][]{Sturm02, 2005ApJ...625L..75H, 2007ApJ...664...71D}.

We find that within the \emph{Spitzer/IRS}  aperture of $\sim3.6$~arcsec
($1-6$~kpc), on average the starburst component contributes $\sim$15~per cent
($\sim$3~per cent PAH and 12~per cent stellar) and the AGN component 85~per cent
to the integrated luminosity of the system in the MIR.  These results
are consistent with the fact that the nuclear MIR emission reported in
the present work is mostly dominated by the AGN (Section
\ref{cc_image}).

\section{Dusty torus modeling}
\label{results2}
\subsection{{\sc clumpy} models}

\begin{figure}
\center
\includegraphics[scale=0.6]{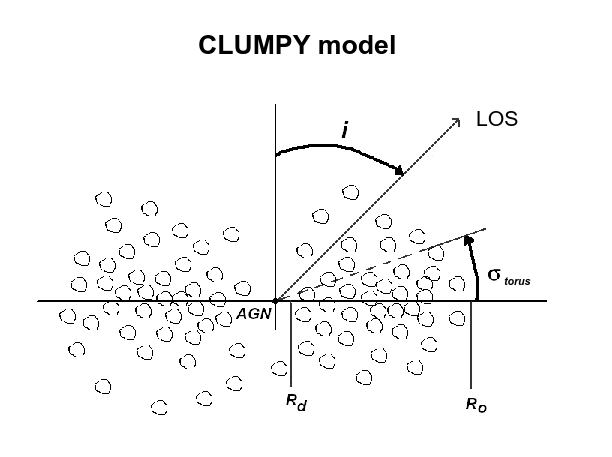} 
\caption{\label{fig0}{\sc clumpy} dusty torus geometry as described in Nenkova et al. (2008a, b). The radial extent of the torus is defined by the ratio between the outer radius $R_{o}$ and the dust sublimation radius $R_{d}$, $Y=R_{o}/R_{d}$, with $Y$ a free parameter. All clouds are assumed to have the same 
optical depth $\tau_{\nu}$. The angular width is given by $\sigma$, and  $N_{LOS}$ is the mean number of 
clouds found along a line of sight at angle  $i$.}
\end{figure}

In this section we adopt 
the {\sc CLUMPY} models of 
\citet{2002ApJ...570L...9N, 2008ApJ...685..147N, 2008ApJ...685..160N} as a
description of the  distribution of clouds that form the dusty torus. 
Within this framework the distribution of clouds is described by a set of 6 
free parameters (see Figure \ref{fig0}). 
The clouds that 
surround
the central engine have the same optical depth, $\tau_{V}$.
The inner radius of the cloud distribution 
is defined by the dust sublimation temperature
($T_{sub}\approx1500$ K for silicates), with $R_{d}=0.4$ (1500 K
$T^{-1}_{sub}$)$^{2.6}(L/10^{45}$erg s$^{-1})^{0.5}$ pc, while the
radial extent $Y$ is defined as the ratio between the
outer ($R_{o}$) and inner radius ($R_{d}$). The radial distribution of
clouds is parameterized as $r^{-q}$, where $q$ is a free parameter. Additionally, there are other three free parameters
that describe the geometry of the torus: the viewing angle
$i$, the angular size $\sigma_{torus}$ and the average number of
clouds along a radial equatorial line $N_{0}$, 
which can be used to calculate the number of clouds along the line of sight 
(LOS) as $N_{LOS}(i)=N_{0}e^{(-(i-90)^{2}/\sigma_{torus}^{2})}$. According to this description the classification of an 
AGN as type 1 or type 2 does not depend only on the viewing angle $i$ but the probability that AGN photons be able to escape through the torus without 
being absorbed 
by an optical thick cloud along the LOS. This is called the escape probability of
AGN-produced photons, $P_{esc}\simeq e^{-N_{LOS}}$. Therefore, it is possible to obtain a type 1 AGN even at viewing angles close to the equatorial plane \citep{2002ApJ...570L...9N, 2008ApJ...685..147N, 2008ApJ...685..160N}.

The emission of {\sc CLUMPY} clouds is the angle-averaged
  emission of all slab orientations. For the models we
  are using, $\tau_{V}$ corresponds to the optical depth along the
  normal of the slab. There is a new version of {\sc CLUMPY}
  models\footnote{\url{www.clumpy.org}} which uses spherical clouds
  with 3-dimensional radiative transfer.

  In order to be able to compare our results with the previous
  modelling performed in lower luminosity AGN 
\citep[e.g.,][]{2011ApJ...736...82A, Alonso16b,
    2013A&A...553A..35G, Gonzalez16, 2015ApJ...803...57I,Martinez15,
    Mateos16, 2011ApJ...731...92R, Fuller16, Garcia-Burillo16}, we chose to use 
  the 2008 {\sc CLUMPY} models \citep{2002ApJ...570L...9N,
    2008ApJ...685..147N, 2008ApJ...685..160N}. 

We model the QSOs with the {\sc BayesClumpy} tool
\citep{2009ApJ...696.2075A}, which uses Bayesian inference to fit the
observed nuclear IR SEDs and MIR spectroscopy of AGN. The output is
the probability distribution of the six free parameters of the model
($\sigma_{torus}$, $Y$, $N_{0}$, $q$, $i$ and $\tau_{V}$) assuming
flat prior information.  {\sc BayesClumpy} allows to fix (or fit) the
redshift and the vertical shift too, i.e. the flux scaling factor of
the model spectrum.  In addition, it is also possible to include
foreground extinction, parameterized by $A_{V}$, when we are modeling
type 2 AGN \citep[see][]{2011ApJ...736...82A}. For type 1 AGN it is
possible to add the AGN (accretion disk) emission in the form of a
power law contribution as described by \cite[][]{
  2002ApJ...570L...9N}. The method uses a Metropolis-Hastings Markov
Chain Monte Carlo (MCMC) sampling technique to determine the posterior
distributions of the free {\sc clumpy} torus model parameters. These
are used in turn to calculate the escape probability distribution ($P_{esc}$) and the
distribution of the geometrical covering factor, $f_{2}=1-
\int_{0}^{\pi/2}P_{esc}(\beta)\cos(\beta){\rm d\beta} $
\citep[e.g.,][]{2011ApJ...736...82A}. For more details on {\sc
  BayesClumpy} see \citet{2014MNRAS.439.3847R}.

\begin{table}
	\begin{minipage}{0.5\textwidth}
		\caption{Parameters of the dusty {\sc clumpy} torus model and assumed range in the prior distributions. 
Columns 2 to 7: angular width, radial extend, number of clouds along the equatorial line, index of the radial density profile, viewing angle and 
			optical depth.
\label{clumpy_tab} } 
				\scalebox{0.85}{
\begin{tabular}{lcccccc} 
\hline
\hline
Parameter & $\sigma_{torus}$ & $Y$ & $N_{0}$ & $q$ & $i$ & $\tau_{V}$ \\
                   & (deg)         &       &      &   &  (deg)  & \\
\hline
Range & $15-70$& $5-100$ & $1-15$ & $0-3$ & $0-90$ & $5-150$\\
				\hline
			\end{tabular}
			}
			 		 
		\end{minipage}				
\end{table}

In order to fit the models, we start by using the unresolved NIR and
MIR starburst-subtracted spectrum between $\sim5-15$ $\mu$m for all
QSOs except PG~1211+143, PG~1351+640, PG~1501+106, and
PG~0050+124. The MIR emission of these three QSOs appears to be
completely dominated by dust heated by the AGN (see Section
\ref{results1}), with negligible starburst contributions. In these
cases we use the IRS spectrum (observed wavelength between $\sim5-30$
$\mu$m) extracted from the database. For three QSOs (PG~0050+124,
PG~1211+143, and PG~1501+106) the spectral range between $\sim5-8$
$\mu$m was excluded (as discussed in Section \ref{individual_qso} and
\ref{discussion}).

In the case of PG~0050+124 the flux density of the unresolved
component at $8.7\mu$m is fainter than the corresponding flux density of the spectrum (see Section
\ref{spec} and Figure in Appendix$^{{\color{blue}4}}$), suggesting the presence of an extended
component in the spectrum larger than the expected uncertainties in flux calibration. Hence we scale the \emph{Spitzer/IRS} spectrum to the $8.7\mu$m photometry data 
point. For this QSO we use the spectral range between $\sim8-20$ $\mu$m, where the emission is strongly dominated by the dusty torus. For MR~2251-178 we used the GTC/CC spectrum.

We remove the emission lines ( [S IV] 10.4 $\mu$m, [Ne II] 12.81 $\mu$m, [Ne V] 14.32 $\mu$m and [Ne III]
15.56 $\mu$m) from the
spectra before fitting the
models, since the emission lines are not modeled.  
We adopt non-informative priors as flat distributions within
the range of values shown in Table \ref{clumpy_tab} for the six free
parameters, and vertical shift values between $-4$ and 4. For all QSOs
we include the direct-light power-law AGN component and do not include a screen of
extinction.

\begin{figure*}
 \includegraphics[scale=0.4]{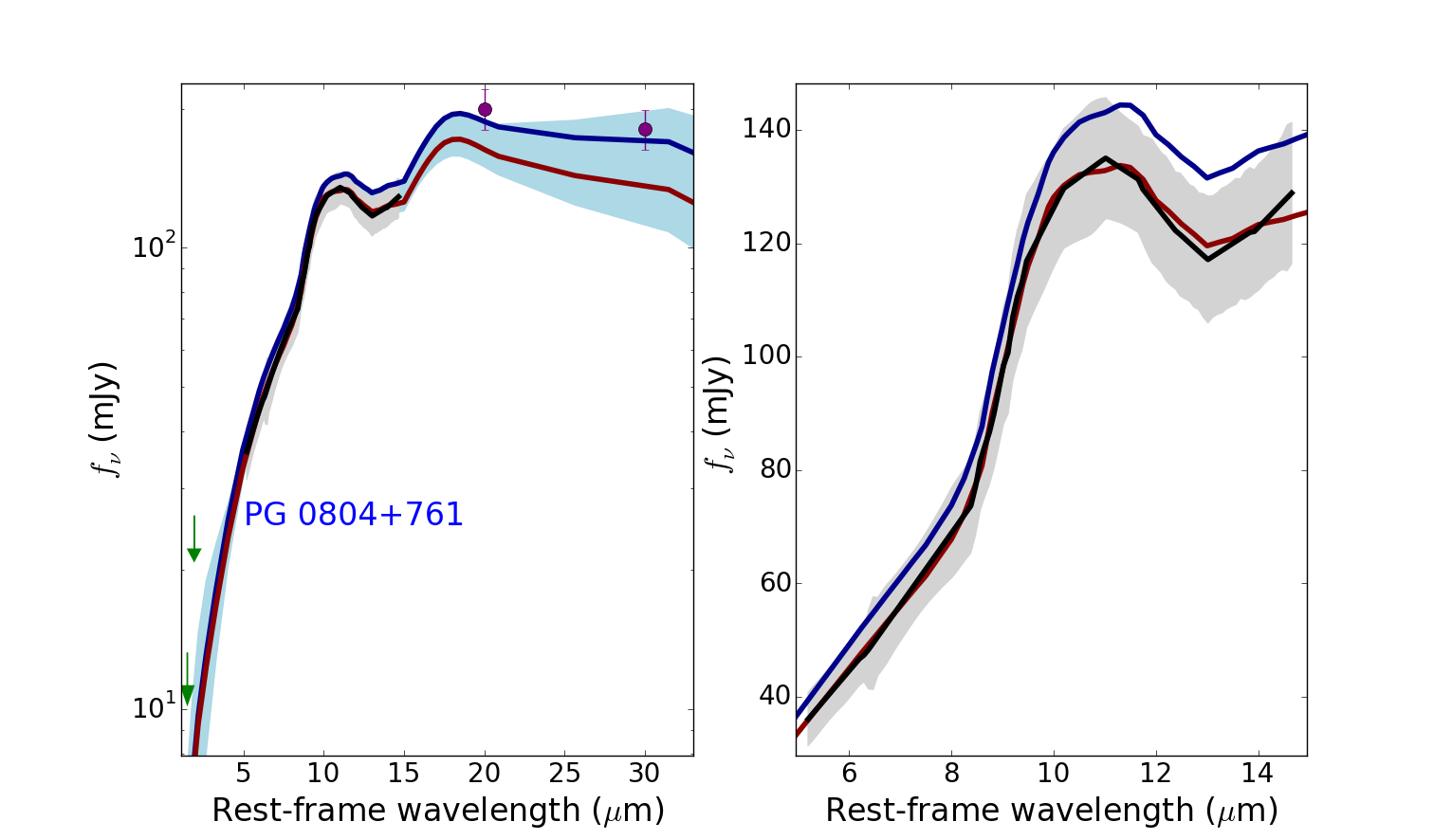} 
	\caption{SED of PG~0804+761.  {\bf Left panel:} 
          photometry for the unresolved component (dots and arrows) and
          starburst-subtracted \emph{Spitzer/IRS}  spectrum (black line). The
          purple dots at 20 and 30$\mu$m are derived from the 
          the starburst-subtracted
          spectrum extrapolating the PAH component obtained from
          the decomposition analysis. We did not to use these data points for the
          modeling. The blue solid line and blue shaded region represent the best
          {\sc clumpy} torus model and the range of models within 68 per
          cent uncertainty in the best-fitted parameters, respectivelly. The
          red solid line is the MAP model. {\bf Right}: enlarged
          view of the best-fit models around the 9.7$\mu$m silicate feature. See Appendix$^{{\color{blue}4}}$ for the rest of the sample.
\label{clump4}}
\end{figure*}

\begin{figure*}
	\includegraphics[scale=0.4]{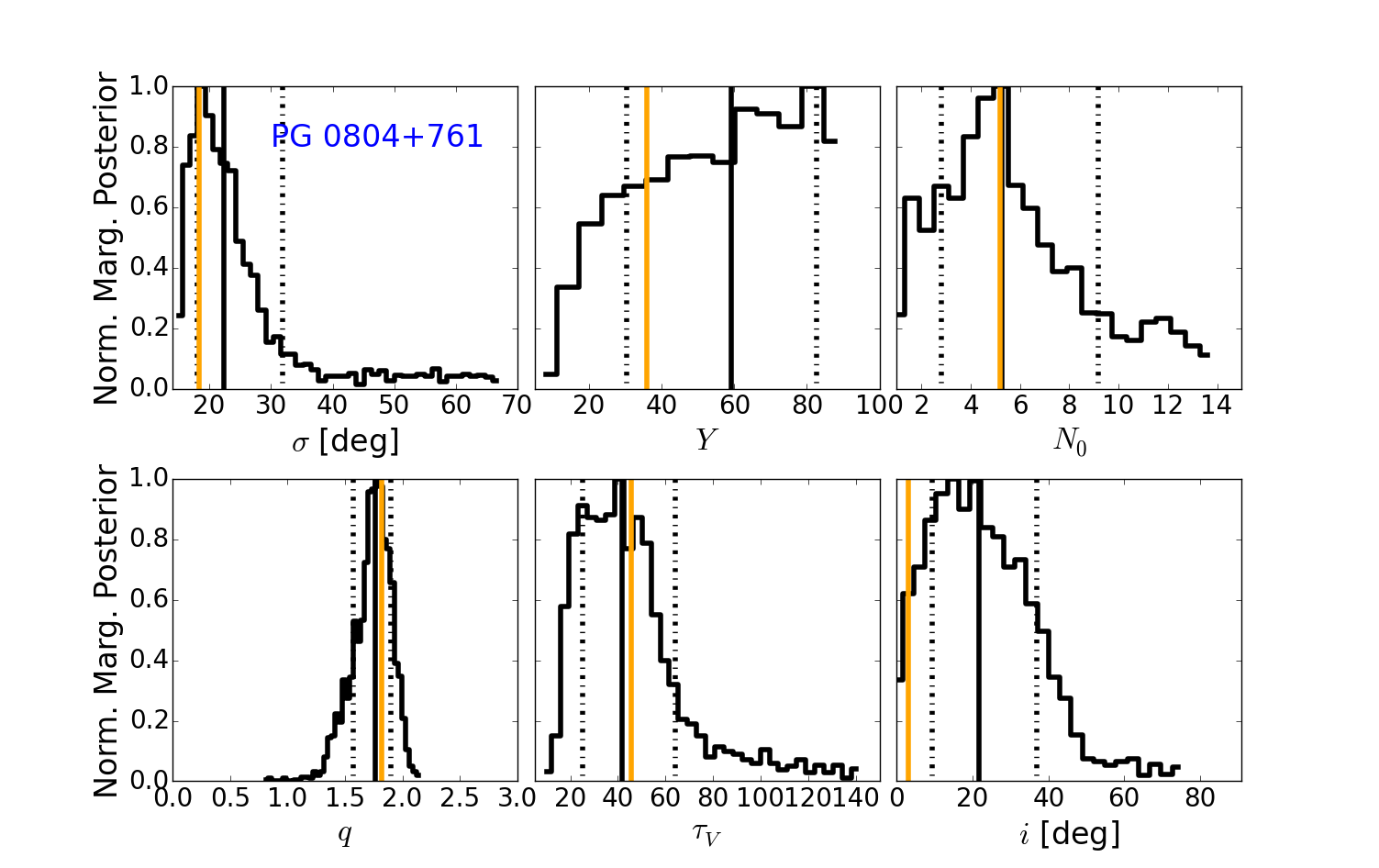} 
	\caption{Posterior probability distributions of the {\sc
            clumpy} parameters for PG~0804+761 in solid black lines. The
          vertical solid black lines mark the median of the
          distributions, the dotted black lines represent the 68~per
          cent confidence intervals, whereas the solid orange line
          marks the MAP values. Posterior probability distribution for the full sample can be found in Appendix$^{{\color{blue}4}}$.
\label{post4}}
\end{figure*}

\begin{table*}

	\begin{minipage}{1.\textwidth}
		\caption{Main results of the dusty {\sc clumpy} torus
                  model fitting. Column 1 gives the name
                  of the QSO; columns 2, 4, 6, 8, 10 and 12 the
                  median and the 68~per cent uncertainty level of the 
                  posterior probability distributions, while columns 3,
                  5, 7, 9, 11 y 13 give the maximum-a-posterior (MAP) model 
                  parameters, and
                  columns 14 and 15 the reduced $\chi_\nu^2$ of the
                  median and MAP models,
                  respectively.\label{clumpy_par} }
                  \hspace{-1cm}
                \scalebox{0.95}{
 \begin{tabular}{l|cccccccccccccc}
                    \hline\\ Name & \multicolumn{2}{c}{$\sigma_{torus}$}  
                    & \multicolumn{2}{c}{$Y$} & \multicolumn{2}{c}{$N_{0}$}& \multicolumn{2}{c}{$q$} &
                    \multicolumn{2}{c}{$\tau_{V}$}& \multicolumn{2}{c}{$i$}&
                    $\chi^{2}_{\nu}$ & $\chi^{2}_{\nu}$ \\
                     & Median & MAP
                    & Median& MAP & Median & MAP & Median& MAP &
                    Median& MAP & Median & MAP &
                     &  \\                    
                    				                &      (deg.)     &  (deg.)            &          &       &       &     &       &       &       &     &  (deg.)  &   (deg.) &(median)&  (MAP) \\
                    \\ \hline \hline PG 0003+199 & $31^{+20}_{-9}$ &
                    43 & $54^{+27}_{-27}$ & 82 & $6^{+3}_{-3}$ & 3 &
                    2.$5^{+0.3}_{-0.4}$ & 1.9 & $77^{+25}_{-21}$ & 67
                    & $22^{+20}_{-13}$ & 65 & 2.15 & 0.12 \\
				
				PG 0007+106	&	$24^{+4}_{-3}$	&	24	&	$58^{+24}_{-27}$	&	79	&	$6^{+2}_{-3}$	&	3	&	2.$6^{+0.2}_{-0.3}$	&	2.1	&	$53^{+12}_{-11}$	&	53	&	$64^{+14}_{-10}$	&	80	&	0.15	&	0.09	\\
				
				PG 0050+124	&	$16^{+1}_{-0}$	&	15	&	$67^{+15}_{-12}$	&	58	&	$3^{+1}_{-0}$	&	3	&	1.5$6^{+0.04}_{-0.03}$	&	1.5	&	$54^{+4}_{-4}$	&	54	&	$79^{+3}_{-3}$	&	80	&	1.37	&	1.25	\\
				
				PG 0804+761	&	$22^{+10}_{-4}$	&	18	&	$59^{+23}_{-29}$	&	36	&	$5^{+4}_{-2}$	&	5	&	1.$8^{+0.1}_{-0.2}$	&	1.8	&	$42^{+22}_{-17}$	&	46	&	$22^{+15}_{-12}$	&	3	&	1.37	&	0.03	\\
				
				PG 0844+349	&	$16^{+1}_{-1}$	&	15	&	$77^{+15}_{-30}$	&	96	&	$2^{+0}_{-0}$	&	1	&	1.$3^{+0.2}_{-0.1}$	&	1.3	&	$77^{+25}_{-15}$	&	66	&	$88^{+1}_{-3}$	&	90	&	0.31	&	0.21	\\
				
				PG 0923+129	&	$43^{+12}_{-10}$	&	36	&	$60^{+23}_{-25}$	&	79	&	$10^{+3}_{-3}$	&	13	&	2.$1^{+0.4}_{-0.4}$	&	2	&	$120^{+10}_{-13}$	&	117	&	$14^{+14}_{-8}$	&	1	&	1.60	&	0.30	\\
				PG 1211+143	&	$45^{+12}_{-8}$	&	36	&	$40^{+4}_{-3}$	&	39	&	$2^{+0}_{-0}$	&	2	&	1.4$3^{+0.05}_{-0.05}$	&	1.4	&	$38^{+3}_{-3}$	&	37	&	$80^{+6}_{-10}$	&	88	&	1.27	&	1.24	\\
				PG 1229+204	&	$16^{+1}_{-1}$	&	15	&	$59^{+20}_{-18}$	&	64	&	$12^{+2}_{-3}$	&	14	&	0.5$3^{+0.10}_{-0.07}$	&	0.5	&	$20^{+9}_{-5}$	&	14	&	$73^{+2}_{-3}$	&	75	&	0.94	&	0.33	\\
				PG 1351+640	&	$17^{+3}_{-1}$	&	15	&	$37^{+1}_{-1}$	&	35	&	$8^{+1}_{-1}$	&	9	&	0.$04^{+0.05}_{-0.03}$	&	0.01	&	$29^{+3}_{-3}$	&	28	&	$47^{+4}_{-9}$	&	51	&	2.13	&	1.77	\\
				PG 1411+442	&	$16^{+1}_{-1}$	&	15	&	$58^{+25}_{-28}$	&	74	&	$4^{+1}_{-1}$	&	3	&	2.$6^{+0.2}_{-0.2}$	&	2.5	&	$72^{+10}_{-10}$	&	73	&	$86^{+2}_{-5}$	&	90	&	0.06	&	0.06	\\
				PG 1426+015	&	$19^{+3}_{-2}$	&	19	&	$60^{+23}_{-27}$	&	50	&	$9^{+3}_{-3}$	&	6	&	2.$0^{+0.4}_{-0.4}$	&	2.7	&	$118^{+14}_{-23}$	&	125	&	$64^{+9}_{-8}$	&	88	&	0.80	&	0.62	\\
				PG 1440+356	&	$16^{+1}_{-0}$	&	15	&	$55^{+26}_{-28}$	&	60	&	$3^{+0}_{-0}$	&	3	&	2.$8^{+0.1}_{-0.2}$	&	2.9	&	$70^{+9}_{-9}$	&	76	&	$88^{+1}_{-3}$	&	90	&	0.08	&	0.06	\\
				PG 1448+273	&	$30^{+7}_{-4}$	&	24	&	$57^{+26}_{-26}$	&	60	&	$13^{+1}_{-3}$	&	15	&	1.$7^{+0.3}_{-0.2}$	&	1.5	&	$89^{+13}_{-12}$	&	73	&	$10^{+10}_{-6}$	&	1	&	0.35	&	0.17	\\
				PG 1501+106	&	$58^{+7}_{-8}$	&	52	&	$40^{+32}_{-18}$	&	21	&	$10^{+3}_{-3}$	&	8	&	2.$6^{+0.2}_{-0.2}$	&	2.4	&	$113^{+16}_{-14}$	&	110	&	$40^{+23}_{-21}$	&	56	&	0.42	&	0.36	\\
				PG 1534+580	&	$28^{+15}_{-6}$	&	18	&	6$7^{+20}_{-26}$	&	80	&	$11^{+2}_{-3}$	&	15	&	1.$8^{+0.6}_{-0.3}$	&	1.5	&	$100^{+31}_{-29}$	&	69	&	$12^{+11}_{-7}$	&	2	&	1.54	&	0.35	\\
				PG 1535+547	&	$37^{+18}_{-13}$	&	24	&	$52^{+26}_{-25}$	&	30	&	$4^{+4}_{-1}$	&	3	&	2.$7^{+0.2}_{-0.3}$	&	2.3	&	$80^{+13}_{-14}$	&	66	&	$47^{+19}_{-22}$	&	79	&	0.19	&	0.06	\\
				PG 2130+199	&	$17^{+2}_{-1}$	&	16	&	$64^{+22}_{-25}$	&	94	&	$5^{+1}_{-1}$	&	4	&	2.$0^{+0.3}_{-0.2}$	&	1.8	&	$50^{+8}_{-7}$	&	46	&	$86^{+3}_{-4}$	&	88	&	0.07	&	0.04	\\
				PG 2214+139	&	$46^{+14}_{-18}$	&	29	&	$54^{+25}_{-25}$	&	45	&	$2^{+1}_{-0}$	&	2	&	2.$3^{+0.2}_{-0.2}$	&	2.1	&	$64^{+25}_{-16}$	&	50	&	$52^{+21}_{-27}$	&	85	&	2.67	&	0.19	\\
				Mrk 509	&	$38^{+14}_{-10}$	&	61	&	$59^{+23}_{-26}$	&	85	&	$8^{+4}_{-3}$	&	2	&	1.$9^{+0.6}_{-0.4}$	&	1.2	&	$99^{+22}_{-29}$	&	58	&	$17^{+15}_{-10}$	&	60	&	2.73	&	0.18	\\
				MR 2251-178	&$37^{+16}_{-11}$	& 36& $46^{+29}_{-26}$	& 67& $7^{+4}_{-3}$& 10& 1.$9^{+0.6}_{-0.7}$& 0.7	& $103^{+25}_{-43}$& 70& $28^{+18}_{-16}$	& 1	& 0.94	& 0.63	\\	
				\hline
			\end{tabular}
			}
		\end{minipage}				
\end{table*}

\subsection{Results for individual QSOs}
\label{individual_qso}
From fitting the unresolved NIR emission and MIR AGN spectrum we note
that for the majority of QSOs (12) the spectral range between
$\sim5-8\,\mu\text{m}$ cannot be reproduced by the {\sc clumpy} models
with a set of parameters consistent with a type 1 AGN. In fact, the
inclusion of this range results in a poor fit of the silicate features
at 9.7 $\mu$m. To reproduce this emission, previous works have
included, apart from the torus emission, a hot dust component
\citep[e.g.,][]{2009ApJ...705..298N, Deo11, Mor2011, Mor_Netzer}.

However, for the 20 QSOs in the sample the spectra are well reproduced
by {\sc clumpy} models when we fit in the $8-15$ $\mu$m range (see
Figure~7 for an example and Appendix$^{{\color{blue}4}}$ for the rest of
the sample).  For four objects (PG~2214+139, PG~0050+124, PG~1440+356
and, PG~1411+442) a comparison of this new fitting with the spectra
still shows an excess of NIR unresolved emission in the 5 to 8 $\mu$m
range.

The posterior probability distributions of the {\sc clumpy} torus
model parameters are well constrained for all QSOs except for
MR~2251-178, which was modeled with the narrower wavelength range
spectrum of GTC/CC. In Table \ref{clumpy_par} we list the median,
$1\sigma$ uncertainty and the maximum-a-posterior (MAP) values of the
parameters derived from the SED modeling. In this table we also list
the reduced $\chi^{2}$ estimated from fitting both the median and MAP
models with the MIR starburst-substrated spectrum. 

 We
  find that 50 per cent of QSOs have viewing angles between 50 and 88
  degrees, while the other 50 per cent has viewing angles that range from
  17 to 47 degrees. A similar analysis on Seyfert 1 galaxies found 
  viewing angles between
  50 and 60 degrees with escape probabilities larger than 16 per cent
  \citep[see e.g.,][]{2011ApJ...736...82A}. Nevertheless, in order to
  better constrain the viewing angle of QSOs, we would need at least two
NIR photometry points \citep[][]{2014MNRAS.439.3847R}.

In Table \ref{clumpy_addpar} we list $P_{esc}$, $f_{2}$ and the AGN
bolometric luminosity derived from the {\sc clumpy} modeling, and we
also include the bolometric luminosity estimated from hard X-rays
(2-10 keV), using bolometric corrections of
\citet{2012MNRAS.426.2677R}, for comparison.

\begin{table*}
 	\begin{minipage}{1.\textwidth}
 		\caption{Parameters derived from the 
{\sc clumpy} free parameters: column 1, name of the QSO; column 2, 
escape probability;
column 3, covering factor; and column 4, bolometric
luminosity. Column 5 gives the range
of bolometric luminosities estimated using hard X-ray (2-10 keV)
fluxes from the literature.\label{clumpy_addpar} }
                \centering
 		\begin{tabular}{l|ccrr} 
 				\hline\\ 
 				Name & $P_{esc}$ &		$f_{2}$	&	$L_{bol}$ {\sc clumpy} & $^{*}L_{bol}$ Obs.\\
 				&  (per cent)  &          &     ($\times10^{44}$ erg s$^{-1}$) &  ($\times10^{44}$ erg s$^{-1}$) \\
 				\hline
 				\hline\\
 				PG 0003+199	&	$92^{+8}_{-51}$	&	$0.3^{+0.2}_{-0.1}$	&	$3.0^{+0.4}_{-0.2}$	&  $(3-12)$\\
 				PG 0007+106	&	$13^{+30}_{-8}$	&	$0.20^{+0.1}_{-0.04}$	&	$24^{+8}_{-4}$	&  $(20-90)$\\
 				PG 0050+124	&	$16^{+6}_{-4}$	&	$0.06^{+0.01}_{-0.04}$	&	$40^{+8}_{-8}$	&  $(10-50)$\\
 				PG 0804+761	&	$99^{+1}_{-8}$	&	$0.2^{+0.1}_{-0.1}$	&	$30^{+2}_{-3}$	&  $(50-190)$\\
 				PG 0844+349	&	$22^{+3}_{-3}$&	$0.040^{+0.01}_{-0.004}$	&	$2.2^{+0.3}_{-0.2}$	&  $(10-37)$\\
 				PG 0923+129	&	$62^{+35}_{-44}$&	$0.7^{+0.2}_{-0.2}$	&	$21^{+2}_{-2}$	&  $(6-23)$ \\
 				PG 1211+143	&	$13^{+3}_{-2}$	&	$0.4^{+0.1}_{-0.1}$	&	$59^{+6}_{-9}$	&  $(9-33)$\\
 				PG 1229+204	&	$3^{+7}_{-2}$	&	$0.1^{+0.01}_{-0.01}$	&	$45^{+12}_{-8}$	&  $(5-21)$\\
 				PG 1351+640	&	$99^{+1}_{-1}$	&	$0.10^{+0.04}_{-0.01}$	&	$8.6^{+0.7}_{-0.7}$	&  $(3-11)$\\
 				PG 1411+442	&	$4^{+2}_{-1}$	&	$0.10^{+0.01}_{-0.01}$	&	$59^{+6}_{-9}$	&  $(4-17)$\\
 				PG 1426+015	&	$29^{+45}_{-25}$&$0.14^{+0.02}_{-0.04}$	&   $45^{+12}_{-8}$	&  $(20-80)$\\
 				PG 1440+356	&	$4^{+1}_{-1}$	&$0.070^{+0.01}_{-0.004}$	&	$86^{+7}_{-7}$	&  $(10-40)$\\
 				PG 1448+273	&	$99^{+1}_{-19}$	&	$0.4^{+0.2}_{-0.1}$	&	$5.7^{+0.5}_{-0.5}$	&  $(3-13)$\\
 				PG 1501+106	&	$1^{+7}_{-1}$	&	$0.9^{+0.1}_{-0.1}$	&	$6.1^{+0.8}_{-0.4}$	&  $(9-33)$\\
 				PG 1534+580	&	$99^{+1}_{-35}$	&	$0.3^{+0.3}_{-0.1}$	&	$21^{+3}_{-4}$	&  $(4-16)$\\
 				PG 1535+547	&	$33^{+47}_{-21}$&	$0.4^{+0.2}_{-0.2}$	&	$0.23^{+0.06}_{-0.03}$	&  $(0.4-1.6)$\\
 				PG 2130+099	&	$2^{+1}_{-1}$	&	$0.09^{+0.02}_{-0.01}$	&	$5.9^{+0.7}_{-0.7}$	&  $(6-22)$\\
 				PG 2214+139	&	$42^{+47}_{-14}$      &$0.3^{+0.1}_{-0.2}$&	$14^{+3}_{-2}$	&  $(10-40)$\\
 				MR 2251-178	&$70^{+28}_{-51}$& $0.4^{+0.3}_{-0.2}$		   &	$72^{+18}_{-12}$ & $(50-190)$	\\
 				Mrk 509	&	$84^{+7}_{-1}$	&	$0.5^{+0.2}_{-0.2}$	   &	$76^{+10}_{-10}$	& $(80-320)$  \\
 				\hline
 			\end{tabular} 
 		\end{minipage}				
\end{table*}
	
We find that Mrk~509 has an index of the radial density profile
($q\sim1.9$) and number of clouds along the equatorial ray
($N_{0}\sim8$) consistent with those previously obtained by
\citet{2010A&A...515A..23H}, $\sim1.5$ and $\sim7.5$ respectively,
using the {\sc clumpy} torus models and an extra
contribution of hot dust in the inner region of the torus
\citep{2010A&A...523A..27H}. We note, however, that they did not fit their MIR spectra but rather compared with a subset of models deemed to be appropriate for Seyfert 1 galaxies. 

PG~1211+143 was also previously modeled by \citet{2009ApJ...707.1550N}
with  the {\sc clumpy} models. Three
of the torus parameters ($N_{0}=2-9$, $\sigma_{torus}=15-60^\circ$ e
$i=0-70^\circ$) are consistent within the uncertainties with our results, 
while we find a different index of
the radial density distribution and larger optical depths ($q=0-0.5$
and $\tau_{V}=20-30$). These differences probably arise 
because they fixed the radial extent to $Y=20$ while we allowed this parameter
to vary freely.

 As a sanity check, we also performed the fits for individual QSOs with the 
new CLUMPY models that use 3-D radiative transfer on spherical clouds, 
available in their web page,  and found that the
parameters are the same as with the 2008 models within the derived 1sigma uncertainties.

\section{Discussion}
\subsection{Global torus properties of QSOs}

\begin{table*} 
    \begin{minipage}{1.\textwidth}
		\caption{Median and 68~per cent confidence intervals  
of the global {\sc clumpy} model parameters for QSOs, type 1 Seyfert nuclei and
type 2 Seyfert nuclei (red). $^{*}$The parameter was fitted using a range between 5 and 30, which is different than range used for QSOs (5-100).
\label{clumpy_par_med}} 
		\centering
		\begin{tabular}{c|cccccccc} 
				\hline\\ 
				AGN &$\sigma_{torus}$ (deg.)	&	$Y$	&	$N_{0}$	&	$q$	&	$\tau_{V}$	&	$i$ (deg.)	& $P_{esc}$ (per cent) & $f_{2}$	\\
				\hline			
				\hline
QSO &				$20^{+25}_{-5}$	& $57^{+28}_{-22}$ & $5^{+6}_{-3}$ & $1.9^{+0.8}_{-0.6}$ & $67^{+40}_{-31}$ & $55^{+32}_{-40}$ & $0.2^{+0.8}_{-0.2}$ &$0.2^{+0.3}_{-0.1}$ \\	
Seyfert 1 &				$20^{+36}_{-4}$	& $^{*}20^{+6}_{-5}$ & $13^{+2}_{-3}$ & $1.2^{+0.7}_{-1.0}$ & $100^{+43}_{-26}$ & $22^{+38}_{-6}$ & $0.2^{+0.8}_{-0.1}$ & $0.2^{+0.8}_{-0.1}$\\	
Seyfert 2 &				$61^{+7}_{-11}$	& $^{*}18^{+6}_{-7}$ & $13^{+2}_{-4}$ & $0.5^{+1.6}_{-0.5}$ & $66^{+42}_{-32}$  & $58^{+9}_{-10}$ & $0.005^{+0.1}_{-0.005}$ & $0.9^{+0.1}_{-0.2}$  \\			
				\hline					
			\end{tabular} 
		\end{minipage}				
\end{table*}

\begin{figure*}
\hspace{-1cm}
\includegraphics[scale=0.3]{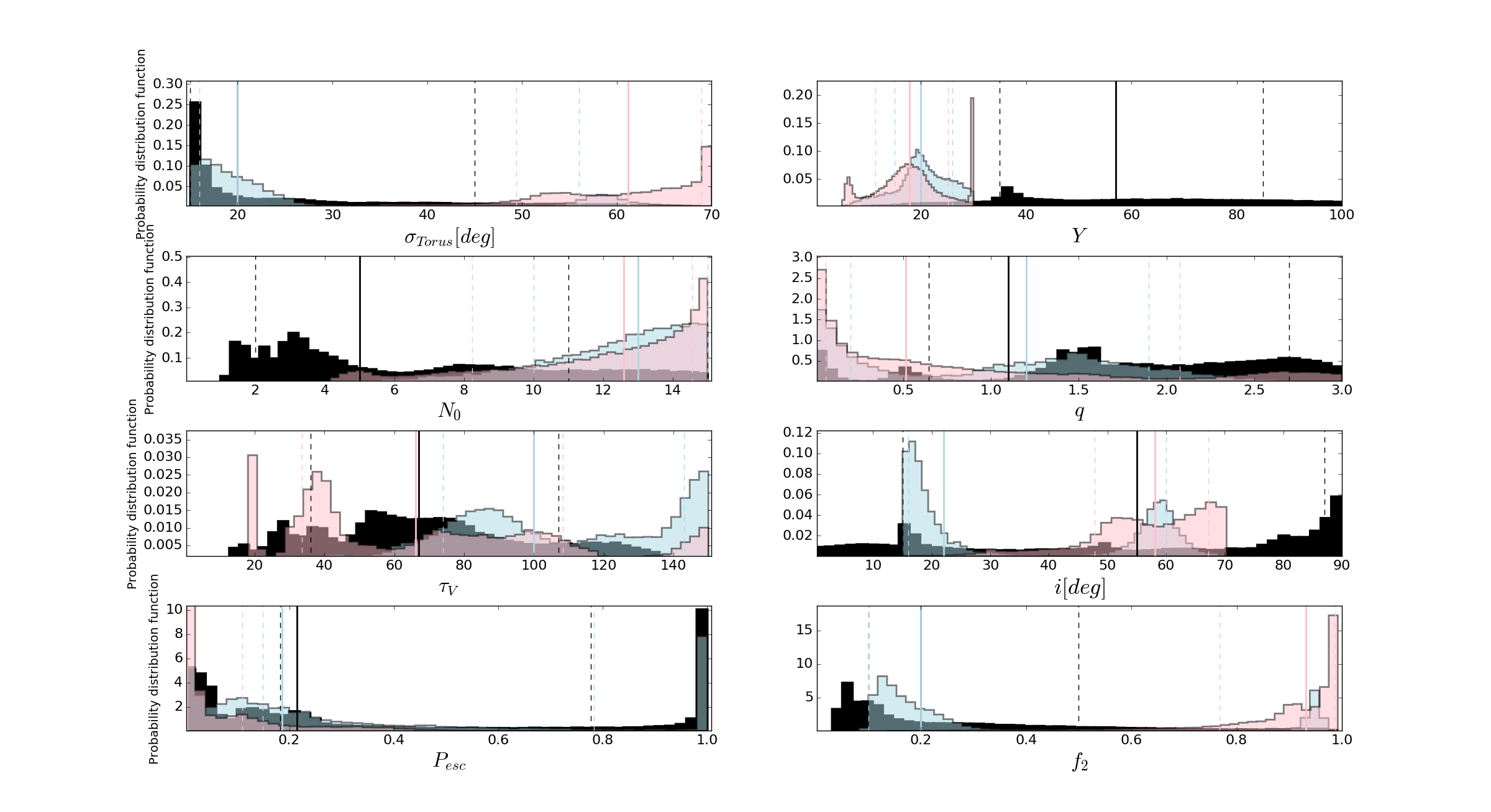} 
\caption{Global probability distributions of the {\sc clumpy} torus parameters for our QSO sample (black) compared with those of type 1 Seyfert nuclei (blue, four objects) and
type 2 Seyfert nuclei (pink) from \citet{2015ApJ...803...57I}.
The solid vertical lines represent the median values of the 
distributions, and the dashed black line shows the 68~per cent confidence interval of the distributions around the median. For Seyfert 1 and 2 the parameter $Y$ was fitted using a range between 5 and 30, which is different than range used for QSOs (5-100).\label{global_par}}
\end{figure*}

In order to study the QSO sample as a whole, we build global
probability distributions for each parameter by bootstrapping on the
parameters returned by the MCMC procedure for each QSO 10,000 times,
and creating a parent distribution of 19 QSOs $\times$10,000 values to
derive the distributions. We exclude MR~2251-178 from the
global analysis since the posterior distributions of its parameters 
are less constrained due to the narrower spectral range used in the analysis.
We build the global probability
distributions for type 1 and 2 Seyfert nuclei in the same manner,
using the individual arrays of values obtained in the analysis
published by \citet{2015ApJ...803...57I} (A. Asensio-Ramos, private communication).  In Figure \ref{global_par}
we show the global probability distributions of QSOs, type 1 and 2
Seyferts. The median values of the distributions and their
68~per cent confidence intervals are listed in Table
\ref{clumpy_par_med}.

We observe that the global probability distributions cover a wide
range of values.  In particular, the escape probability for QSOs,
which depends on the number of clouds along the LOS ($N_{LOS}$), the viewing angle ($i$) and the angular width ($\sigma_{torus}$),
shows a peak below 0.1 (5 objects with $P_{esc}<5$~per cent), a
secondary peak around 0.2 (7 QSOs with $10<P_{esc}<70$~per cent), and
a peak above 0.7 (7 QSOs with $P_{esc}>70$~per cent). Nevertheless, better constrains on the viewing angle ($i$)
  could result in better constrains on the escape probability
  distribution. The geometrical covering factor, which is independent of the viewing angle, however, is well
constrained towards low values (median of $f_{2}\sim0.2$).  Therefore,
QSOs display a wide range of global {\sc clumpy} model properties
($\sigma_{torus}$, $N_{0}$, and $i$) in combinations such that
$P_{esc}$ and $f_{2}$ allow for enough AGN-produced photons to escape
the dusty structure and the broad lines to be seen in direct light,
resulting in a type 1 QSO.

A qualitative comparison of the {\sc clumpy} model parameters for QSOs
with those of Seyfert 1 and 2s shows that the distribution of
number of clouds $N_{0}$ is skewed in QSOs towards lower values than
in Seyfert galaxies, and they are more
concentrated towards the inner regions of the torus (larger $q$
values). The optical thickness of the clouds ($\tau_\nu$) is lower than in
Seyfert 1s and comparable to Seyfert 2s.
The values of $\sigma_{torus}$ in QSOs and Seyfert 1s,
however, are more similar, albeit with different parent distributions.
The $Y$ parameter peaks at lower values in Seyfert
galaxies than in QSOs, although it is not possible to strictly compare
the distributions since \citet{2015ApJ...803...57I} constrain its range
between 0 and 30, while we allowed it to vary between 0 to
100, and indeed, we note that the global distribution of $Y$ for Seyfert
1s is  truncated at 30. The viewing angle is $\sim \times 1.5$ larger in QSOs than in Seyfert 1s, but
the median values of $P_{esc}$ and $f_{2}$ are similar in QSOs
and Seyfert 1s. 

To quantitatively compare the probability distributions of the
parameters, we use the two-sample Kolmogorov-Smirnov (KS) test to
determine the probability of rejecting the null hypothesis that two
samples are drawn from the same parent population. For all free
parameters ($\sigma_{torus}$, $Y$, $N_{0}$, $q$, $i$, $\tau_{V}$) of Seyfert 1 and 2 the
null hypothesis can be rejected with a negligible probability
($P< 10^{-100}$). Likewise $P_{esc}$ and $f_{2}$ have
statistically different parent distributions, and the null hypothesis
can again be reject with $P< 10^{-100}$. These probabilities
are so small due, in part, to the large number of samples in the
global distributions that map the posterior probability distributions
of individual AGN, but caution is drawn to the fact that the
distributions have been derived from only $\sim$ 10-20 objects for QSOs and Seyfert 2, and four for Seyfert 1. 

Additionally, we use the {\it Mann-Whitney} \citep[MW,][]{Stuart94} test to statistically
measure the similitude between the medians of distributions. Although
for QSOs and Seyfert 1s the medians look similar in $\sigma_{torus}$,
$P_{esc}$ and $f_{2}$ parameters, they are statistically distinct
($P<10^{-7}$), and QSOs do have a tendency for larger escape
probabilities and lower covering factors than Seyfert 1 nuclei.

These results are consistent with fundamental geometrical differences
between high luminosity type 1 AGN (QSOs) and their lower luminosity
counterparts (Seyfert 1s), and between those type~2 AGN.

The low number of clouds along the equatorial ray ($N_{0}$), large
index of the radial distribution ($q$), and lower optical thickness
($\tau_{V}$), suggest that the clouds in the
tori of QSOs might have been partially evaporated and piled away 
by the high radiation field of the QSOs, as
proposed by the receding torus scenario \citep[][]{Lawrence91}. 

\subsection{Dependence of AGN covering factor on AGN luminosity}
\label{discussion} 
 Several works \citep[e.g.,][]{Hasinger08, Ueda03, LaFranca05,
    Gilli07, Fiore08} have found that the fraction of type 2 AGN
  decreases with increasing AGN luminosity, and this result can be
  interpreted as giving support to a model in which the torus recedes
  due to the higher intensity radiation field
  \citep[][]{Lawrence91}. However, these works have used the fraction of absorbed AGN from X-ray or optical broad lines, which could be affected by absorption along the LOS due to dust in the host galaxy or dust-free ionaized gas \citep[see][and references therein]{Rowan-Robinson09}.

   Using a large X-ray and IR selected
  sample of AGN \cite{Rowan-Robinson09} studied the X-ray-infrared
  correlation. They found that their data is well reproduced by a
  model in which the median covering factor decreases from high
  ($log_{10}L_{X}>44.5$) to moderate ($log_{10}L_{X}=42.5-44.5$)
  X-ray luminosities and then increases towards low luminosities
  ($log_{10}L_{X}=42.5$).

\begin{figure}

\includegraphics[scale=0.36]{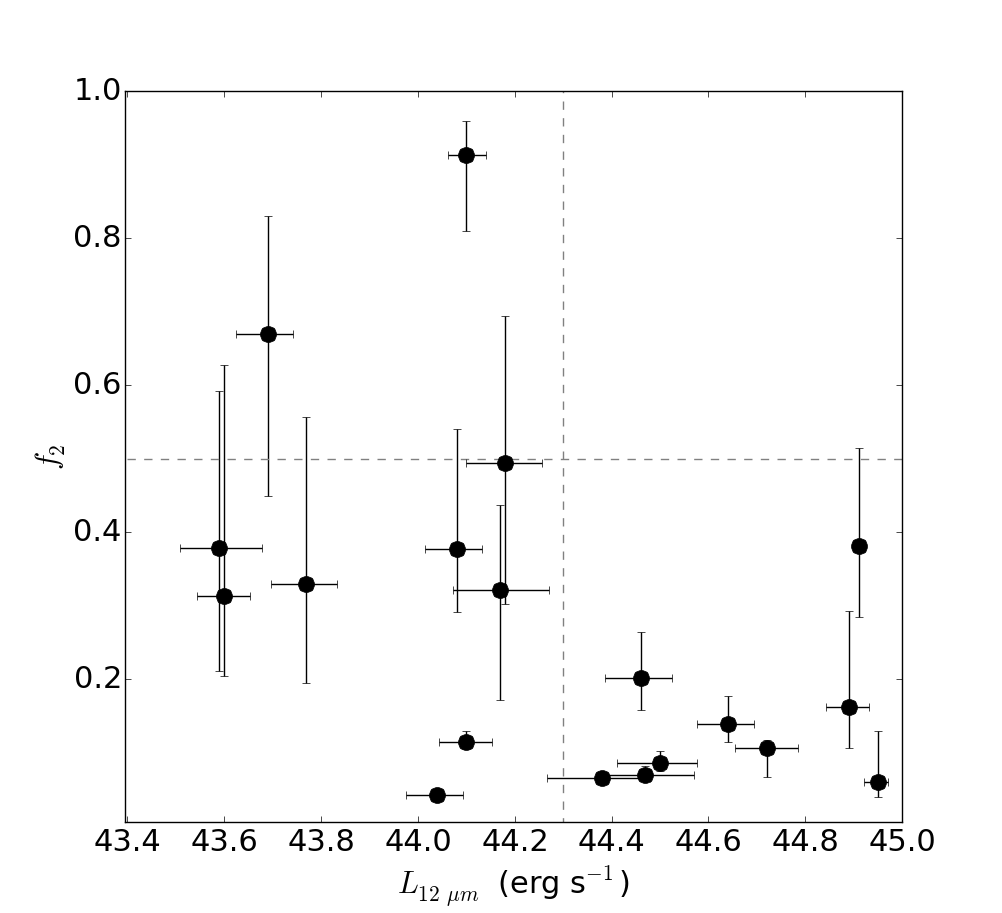} 	
\caption{Geometrical covering factor $f_{2}$ as a function of the IR luminosity at 12 $\mu$m.\label{f2_L12}}		
	\end{figure}

\begin{figure}

\includegraphics[scale=0.35]{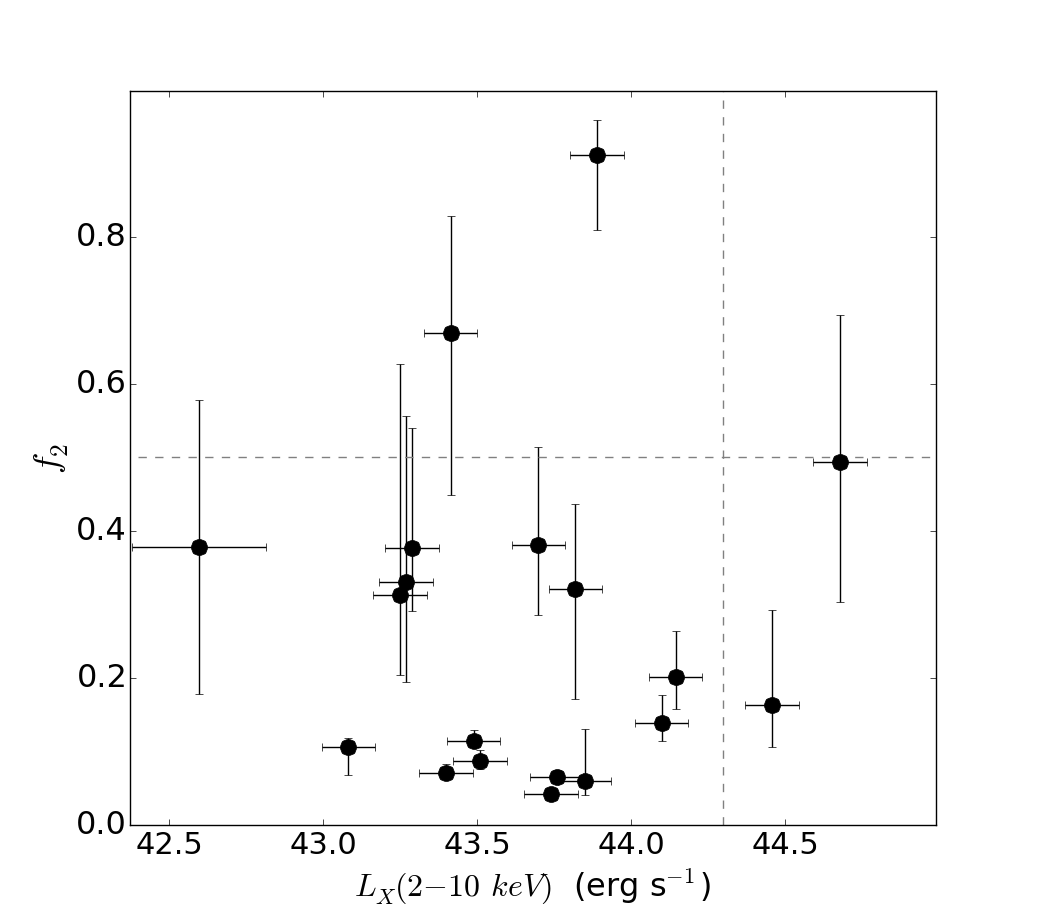} 	
\caption{Geometrical covering factor as a function of the hard (2-10 keV) X-ray luminosty.\label{f2_Lx}}	
	\end{figure}

In our sample, we find that objects with lower ($log_{10}L<44.3$) IR and X-ray
luminosity have a wide range of covering factors, reaching up to the
highest values (Figures \ref{f2_L12} and \ref{f2_Lx}). These 12$\mu$m
luminosities are hence dominated by the dusty torus emission.  In
particular, we note that PG~1501+106, the QSO with the highest $f_2$
value, is among the objects with low IR and X-ray emission. The high
luminosity and high covering factor region of this relationship,
$f_{2}>0.5$ and $L_{12}>44.3$, is devoid of QSOs. Although only 20
objects have been included in this comparison, the absence of QSOs in
this region is significant. If both quantities were to be completely
unrelated, we would expect a scatter plot covering all values, with a
mean of 7.6 objects falling in the $f_{2}>0.5$ and $L_{12}>44.2$
region in the case of $f_{2}$ vs$ L_{12\,\mu m}$ relationship. The
probability of finding in a sample 0 objects when 7.6 are expected is
$P<0.05$~per cent, assuming a Poisson distribution. Hence the
highest-luminosity AGN have a high tendency to have cleaner line of
sights. We note that the two QSOs with $f_{2}>0.5$ (PG~0923+129 and
PG~1501+106) are 1.2 and 1.5 type objects. The median covering
  factor of QSOs is $0.2^{+0.3}_{-0.1}$ and this value is consistent,
  within the uncertainties, with the 35-40~per cent derived by
\citet{Rowan-Robinson08, Rowan-Robinson09} for a sample of QSOs with
  $log_{10}L_{X}$ in the range of  42.5 to 44.5 erg s$^{-1}$.

 Recently, \citep[][]{Mateos16} used a large sample of AGN from the
  Bright Ultra-hard XMM-Newton Survey that includes type 1s, type 2s and
  intermediate AGN. The torus emission of this sample was modeled
  using the {\sc CLUMPY} models of \citep[][]{2008ApJ...685..147N,
    2008ApJ...685..160N}. All free parameters were allowed to vary within 
  the same
  range as we used for our sample of QSOs, except the radial extend ($Y$), 
  that was constrained a priori between
  5 and 30. They found that type 2 AGN present on average higher
  ($>0.5$) geometrical covering factors than type 1 AGN
  ($<0.5$), but they also found that $\sim 20$ per cent of type 1 AGN
  have covering factors larger than 0.5 and that the $\sim 28$ per cent of
  type 2 AGN have covering factors lower than 0.5. These results are
  consistent with ours, since most QSOs in our sample have covering
  factors $f_2<0.5$.

\subsection{Excess NIR and MIR emission: starburst, hot dust or accretion disk?}
Since {\sc clumpy} models only take into account the emission of the dusty torus we have 
done our best to exclude any extended NIR and MIR emission.
However, in section~6 we found that for some  QSOs 
the unresolved NIR emission is not well
reproduced  by {\sc clumpy} models. Among them, PG~1440+356 has the most
prominent PAH emission, observed in both \emph{Spitzer/IRS} and GTC/CC
spectra (see Figure in Appendix$^{{\color{blue}4}}$), and PG~0050+124 presents clear
extended emission at NIR \citep{1999ApJ...512..162S} and MIR
wavelengths \citep[][and this work, see Figure
  in Appendix$^{{\color{blue}4}}$]{2013A&A...558A.149B}. Therefore, it is likely that
in these objects the extended NIR emission has not been properly
accounted to estimate the unresolved emission, or it 
could also be hot dust associated with the NLR \citep[][]{2010A&A...515A..23H, 2009ApJ...705..298N, Hernan-Caballero16, Mateos16}.
 Similar
results have been reported in the modeling of the unresolved NIR SED
of some Seyfert 1s \citep[e.g.,][]{2011ApJ...736...82A, 2015ApJ...803...57I}
suggesting that in type~1 AGN it is more difficult to constrain the emission from dust heated by the AGN. Another possible explanation is that a power law is not adequate to represent the intrinsic AGN emission in the NIR \citep[][]{Mateos16, Hernan-Caballero16}.

\section{Conclusions}
\label{conclusions}

The MIR imaging of our sample of 20 nearby QSOs reveal that 
 the majority of them are
unresolved at $\sim 0.3$~arcsec resolution, corresponding to physical scales of $\lesssim 600$~pc.  We compare the \emph{Spitzer/IRS} and the ground-based 
high-angular resolution spectra and
find that the spectral shapes are similar, and hence adopt the former to 
characterise the AGN component. We find that on average
the AGN contributes 85~per cent of the total MIR emission within the 
\emph{Spitzer/IRS} apertures, while the rest can be attributed to 
starburst emission. These results indicate that
at MIR wavelengths 
the QSO emission is largely dominated by dust heated by the AGN.

We use the unresolved NIR emission and the starburst-subtracted MIR
spectra to constrain the physical and geometrical parameters of 
{\sc clumpy} dusty torus models. Using Bayesian inference we derive the posterior probability
distributions of the six free parameters of the {\sc clumpy} models and build the
global probability distributions of the parameters for the QSO sample.
We find that for most QSOs {\sc clumpy} models reproduce well the AGN emission without the 
inclusion of a hot dust component, as proposed in the literature.  

A statistical analysis reveals that the properties of the dusty torus
are intrinsically different from those of Seyfert 1 and Seyfert 2
nuclei \citep[e.g,][]{2011ApJ...731...92R, 2015ApJ...803...57I}. Nevertheless, in QSOs the combination of the width of torus,
number of clouds and inclination
$\sigma_{torus}$,
$N_{0}$ and $i$ 
results in escape probabilities $P_{esc}\gtrsim5$~per cent
and covering factors $f_{2}\lesssim0.6$, which are consistent with the
optical classification of QSOs as type 1 AGN. Higher luminosity QSOs have the lowest covering factor $f_{2}$.
We conclude that the lower
number of clouds, steeper radial distribution and less optically thick
clouds in QSOs can be interpreted as dusty
structures that have been partly evaporated and piled up by the higher
intensity radiation field in QSOs, as proposed by a receding torus scenario.

\section*{Acknowledgements}
 We thank the referee of this article for insightful comments that allowed us to improve the manuscript. This work has been partly supported by Mexican CONACyT grant CB-2011-01-167291. MM-P acknowledges support by the CONACyT PhD
fellowship program and UNAM-DGAPA postdoctoral fellowship. OG-M acknowledges support by the PAPIIT IA100516 grant. AA-H acknowledges financial support from the Spanish Ministry of Economy and Competitiveness through the Plan Nacional de Astronom\'ia y Astrof\'isica through grant AYA2015-64346-C2-1-P and from CSIC/PIE grant 201650E036. CRA acknowledges the Ramón y Cajal Program of the Spanish Ministry of Economy and Competitiveness through project RYC-2014-15779. KI acknowledges support by JSPS fellowship for young researchers (PD). This work is
based on observations made with the 10.4m GTC located in the Spanish
Observatorio del Roque de Los Muchachos of the Instituto de
Astrof\'isica de Canarias, in the island La Palma. It is also based
partly on observations obtained with the \emph{Spitzer Space
  Observatory}, which is operated by JPL, Caltech, under NASA contract
1407. This research has made use of the NASA/IPAC Extragalactic
Database (NED) which is operated by JPL, Caltech, under contract with
the National Aeronautics and Space Administration. CASSIS is a
product of the Infrared Science Center at Cornell University,
supported by NASA and JPL.







\appendix
\section{Extended emission}
\label{individual} PG~0050+124  has a radial profile which is clearly more
extended than the radial profile of the standard star, as can be seen
in Figure \ref{cc_1}. Therefore, for this object we model the
brightness profile using the 2D algorithm {\sc GALFIT}
\citep{2002AJ....124..294P}. For that we assume the following models:
(1) a S\'ersic and PSF components with all parameters left to vary
freely; (2) a S\'ersic component with all parameters allowed to vary,
except the index of the brightness profile $n$, which we assume as
$n=$1, plus a PSF component with its parameters free; and (3) a
S\'ersic profile with $n=4$ and a PSF component with free parameters. 
We find that the model that best reproduces the MIR
emission of PG~0050+124 with a reduced $\chi^{2}_\nu \sim1$ is the one that
includes a S\'ersic component with $n=4$ plus a PSF. 
The flux in the residual image is
$\sim1.8$~per cent. The parameters of the S\'ersic component and the extended
and unresolved emission are listed in Table \ref{IZw1_flux}. These
parameters are consistent with those found by
\cite{2006ASPC..357..231V} at $H$ band using HST/NICMOS images.
The uncertainties reported in Table \ref{IZw1_flux} are the standard
deviation of the values given by all models. The unresolved plus
S\'ersic flux of PG~0050+124 measured with {\sc GALFIT} is similar,
within the uncertainties, to that measured directly on the image
inside an aperture radius of 1 arcsec.  The residual image (see
 Fig. \ref{cc_2}) reveals the possible presence of
a ring in PG~0050+124.  Indeed,
\citet{1998ApJ...500..147S} detected a ring using observations of
$^{12}$CO(2-1) and $^{13}$CO(1-0) lines with the four antennas of the
Plateau de Bure interferometer from IRAM. However, the size of the
apparent ring ($\sim1$ arcsec of diameter) observed in the residual
image of CC does not fit the size of the ring (1.6 arcsec of diameter)
reported by \citet{1998ApJ...500..147S}. On the other hand,
  this extended emission could alternatively be 
related to the
  silicate extended region proposed by \cite{Schweitzer08}. 

\begin{table} 
\begin{minipage}{0.5\textwidth}
	\caption{Results from the {\sc GALFIT} modeling using a PSF + S\'ersic
		profile model for PG~0050+124 in the Si2 band ($8.7\, \mu$m).
Column 1, unresolved nuclear emission; column 2, 
integrated flux of the S\'ersic component; 
column 3, index of the S\'ersic profile ($^*$for the 1.60 
and $2.22\,\mu$m model the index was  fixed); 
column 4, effective radius of the S\'ersic component; column 5, 
axis ratio of the 
S\'ersic component; column 6, PA of the major axis 
of the S\'ersic component measured East to North.
} 
	\begin{tabular}{cccccc} 
		\hline
		$f_{\rm unresol}$ & $f_{\rm Sersic}$ & $n$ & $r_{\rm eff}$& $a/b$ & PA\\
		 (mJy) & (mJy) &   &(pc) & & (degree) \\
		\hline
		$57\pm1$&$173\pm1$&  4$^*$& $1520\pm210$ & $0.99\pm0.03$ & $35\pm6$  \\
		
		\hline
	\end{tabular}
	\label{IZw1_flux} 
	\\
\end{minipage}
\end{table}


\bsp	
\label{lastpage}
\end{document}